\documentclass[a4paper,aps,twocolumn,nobibnotes,rmp,superscriptaddress,longbibliography]{revtex4-1}
\usepackage{color}
\usepackage[table,xcdraw]{xcolor}
\usepackage{tabularx}
\usepackage{graphicx}
\usepackage{amssymb}
\usepackage{amsmath}
\usepackage{bm}
\usepackage{verbatim}
\usepackage[utf8]{inputenc}
\usepackage{bbm}

\usepackage{xcolor}
\definecolor{dark-gray}{rgb}{.35,.55,.55}
\definecolor{dark-blue}{rgb}{.0,.0,.6}
\usepackage[colorlinks=true,linkcolor=dark-blue,citecolor=dark-blue,urlcolor=dark-blue]{hyperref}

\usepackage[mathscr,mathcal]{euscript}

\newcommand{\hil}{\mathcal{H}}
\newcommand{\mean}[1]{\ensuremath{\langle#1\rangle}}
\newcommand{\ket}[1]{\ensuremath{|#1\rangle}}
\newcommand{\bra}[1]{\ensuremath{\langle#1|}}
\newcommand{\braket}[2]{\ensuremath{\langle #1|#2\rangle}}
\newcommand{\ketbra}[2]{\ensuremath{| #1 \rangle \langle #2 |}}
\newcommand{\tr}{{\rm tr}}

\newcommand{\Tr}{\operatorname{tr}}

\newcommand{\be}{\begin{eqnarray}}
\newcommand{\ee}{\end{eqnarray}}

\usepackage{tikz}

\newcommand{\abs}[1]{{| #1 |}}

\begin{document}

\title{Incompatible measurements in quantum information science}
\author{Otfried G\"uhne}
\affiliation{Naturwissenschaftlich-Technische 
Fakultät, Universität Siegen, Walter-Flex-Straße 3, 57068 Siegen, Germany}
\author{Erkka Haapasalo}
\affiliation{Centre for Quantum Technologies, National University of Singapore, Science Drive 2, Block S15-03-18, Singapore 117543}
\author{Tristan Kraft}
\affiliation{Institute
for Theoretical Physics, University of Innsbruck, Technikerstraße 21A, 6020 Innsbruck, 
Austria}
\affiliation{Naturwissenschaftlich-Technische 
Fakultät, Universität Siegen, Walter-Flex-Straße 3, 57068 Siegen, Germany}
\author{Juha-Pekka Pellonp\"a\"a}
\affiliation{Department of Physics and Astronomy, University of Turku, FI-20014 Turun yliopisto, Finland}
\author{Roope Uola}
\affiliation{D\'{e}partement de Physique Appliqu\'{e}e, Universit\'{e}  de Gen\`{e}ve, CH-1211 Gen\`{e}ve, Switzerland}
\date{\today}

\begin{abstract}
Some measurements in quantum mechanics disturb each other. This has puzzled physicists since the formulation of the theory, but only in recent decades has the incompatibility of measurements been analyzed in depth and detail, using the notion of joint measurability of generalized measurements. In this Colloquium joint measurability and incompatibility are reviewed from the perspective of quantum information science. The Colloquium starts by discussing the basic definitions and concepts. An overview on applications of incompatibility, such as in measurement uncertainty relations, the characterization of quantum correlations, or information processing tasks like quantum state discrimination, is then presented. Finally, emerging directions of research, such as a resource theory of incompatibility as well as other concepts to grasp the nature of measurements in quantum mechanics, are discussed.
\end{abstract}

\maketitle

\tableofcontents

\section{Introduction}
\label{sec-intro}
Measurements in quantum mechanics are different than their classical 
counterparts. From today's perspective this statement may sometimes 
seem to be a truism or platitude, but when quantum theory was developed 
the notion of measurements and their relation to physical quantities was 
indeed a major roadblock on the way to a better understanding. In 1925, 
Werner Heisenberg noted that the product of physical quantities in the 
theory of atoms may depend on their order \cite{heisenberg1925}. Directly 
thereafter, Max Born and Pascual Jordan pointed out that the fundamental 
reason for this is that physical quantities in quantum mechanics are 
described by matrices \cite{bornjordan1925}. Matrix calculus was not 
common knowledge to physicists those times, so Born and Jordan found it 
important to point out directly at the beginning of 
their paper that for two general matrices $A$ and $B$ 
\begin{equation}
AB \neq BA
\end{equation}
holds. But what is the physical relevance of this non-commutativity?

In the following years, the fact that two observables do not share common
eigenstates attracted attention in the form of uncertainty relations 
\cite{heisenberg1927, kennard1927, robertson1929, robertson1934}. 
Here the non-commutativity directly plays a role, such as in the Robertson relation
\begin{equation}
\Delta (A) \Delta(B) \geq \frac{1}{2} \vert \bra{\psi} [A,B] \ket{\psi} \vert,
\label{eq-robertson}
\end{equation}
where $\Delta (A)$ denotes the standard deviation of the observable $A$, and
$[A,B] = AB-BA$ is the commutator. Due to such relations, non-commutativity
of observables is sometimes seen as a key phenomenon in quantum mechanics, 
containing already most of the mysteries of quantum measurements. 

It turned out, however, that the notion of observables or Hermitian matrices
is much too narrow to describe all measurements in quantum mechanics 
\cite{Ludwigbook, Daviesbook, Helstrombook, Holevobook, Prugoveckibook, busch16}.
Indeed, the textbook notion of projective measurements can be extended to 
positive-operator-valued measures, briefly, POVMs (a short historical 
review is given by \citet{ali09}). POVMs can have more outcomes and may be seen as 
measurements carried out with the help of an additional quantum system. They 
are at the core of the modern formulation of operational quantum mechanics 
and provide an advantage in fundamental protocols of quantum physics, 
such as the discrimination of quantum states. POVMs are the most general 
description of the outcome statistics of measurements, but if the 
post-measurement state is taken into account, one needs to further generalize 
them and consider so-called quantum instruments, introduced by \citet{davies1970}. 

But what is the extension of the notion of non-commutativity to POVMs? Here, several
notions have been introduced, but their relation was often not clear and a direct
physical interpretation was missing. In recent years, however, the situation has
changed. The notion of {\it joint measurability} of POVMs has turned out to be
fundamentally related to several other phenomena in quantum mechanics and quantum
information theory. Joint measurability is related to measurement uncertainty
relations as well as preparation non-contextuality. Moreover, {\it incompatibility} 
(i.e., the absence of joint measurability) is essential for the creation and 
exploitation of quantum correlations, e.g., in the form of quantum steering. 

In this article, we give an overview on joint measurability from the perspective
of quantum information theory. Starting from the basic definitions and properties 
of joint measurability and related concepts, we discuss their applications, e.g., 
in Bell nonlocality or protocols in quantum information processing. Our aim is to 
present these concepts in a simple language, 
in order to serve as an introduction for researchers from different backgrounds. 

We note that 
several excellent works exist, which cover parts of the theory presented in our article. 
For instance, incompatibility from an operational point of view was discussed 
by \citet{heinosaari16b} and quantum measurement theory from a mathematical 
perspective was in depth developed by \citet{busch16}. In addition, 
joint measurability is connected to several topics of quantum information theory,
and interested readers can find several detailed overview articles about them. These
topics include quantum correlations like Bell nonlocality \cite{brunner14} or  quantum 
steering \cite{cavalcanti17, uola20review}, and phenomena and applications like
uncertainty relations \cite{buschrmp2014}, quantum contextuality \cite{liang2011, budroni21}, and 
quantum state discrimination \cite{Barnett09}. 

In detail, this article is structured as follows. In Section~\ref{sec-concepts}
we motivate and explain basic concepts to describe measurements. This includes 
the central notion of POVMs and their joint measurability, as well as concepts 
like instruments and the disturbance of measurements. Section~\ref{sec-witness}
discusses important results on joint measurability. We start with analytical results
for qubit systems, and discuss then measures of incompatibility including their numerical evaluation via semidefinite programming, and constructive methods to obtain joint measurements.
Section~\ref{sec-correlations} connects joint measurability with different concepts
in quantum information processing. We describe intimate connections to various
forms of quantum correlations, to foundational effects such as contextuality and 
macrorealism, and to information processing tasks like state discrimination or 
random access codes. Finally, Section~\ref{sec-applications} collects various extensions
of the concepts discussed so far, including resource theory aspects, other notions for 
accessing the nonclassical behaviour of quantum measurements, and the incompatibility of
quantum channels.

\section{Concepts}
\label{sec-concepts}

Throughout this review, we use the measurement theoretical formulation of quantum measurements, see e.g.\ \cite{busch16}.
In this formulation, Hermitian operators are generalised to positive operator valued measures, 
and the state updates caused by measurements are described by quantum instruments. The latter are 
objects generalising the projection postulate of the Hermitian formulation. This corresponds 
to the most general model for quantum measurements and, importantly to this review, manages 
to describe the different operational formulations of measurement incompatibility.

\subsection{Measurements and instruments}
\label{subsec:measurements}
A positive operator valued measure (POVM) is a collection of positive semi-definite matrices $\{A_a\}$ which normalises to the identity operator, i.e.\ $\sum_a A_a=\openone$. The positivity and normalisation requirements correspond to the requirement on the related measurement outcome statistics to form a probability distribution. In a quantum state $\varrho$, i.e., a positive unit-trace operator, these probabilities are given by $p(a|\varrho)=\text{tr}[\varrho A_a]$, where $a$ is the outcome. Whenever $A_a$ is a projection for all $a$, i.e.\ $A_a^2=A_a$, the POVM is said to be sharp or a projection valued measure (PVM for short). The special case of PVMs is in one-to-one correspondence with the Hermitian formulation by the spectral theorem, i.e.\ any Hermitian operator is of the form
$\sum_a a P_a$ for some unique PVM $\{P_a\}$. Although POVMs are more general than PVMs, any POVM can be seen as a PVM on a larger system through the Naimark dilation, see Section~\ref{subsubsec:Naimark}. 

When we describe the entire measurement process, we have to take into account how the state changes conditioned on registering an outcome. If an outcome $a$ is obtained, the non-normalized post-measurement state is $\sigma_a$, and we assume that the map $\varrho\mapsto\sigma_a$ is a linear (or rather affine) completely positive map and the sum $\sum_a\sigma_a$ is a quantum state. Thus, a measurement is associated to an instrument $\{\mathcal{I}_a\}$ which is a collection of linear completely positive maps such that the sum $\sum_a\mathcal{I}_a$ is a completely positive trace-preserving (CPTP) map, i.e., a quantum channel. It is easy to see that the projection postulate $\varrho\mapsto P_a\varrho P_a$ for a PVM $\{P_a\}$ is an instance of a quantum instrument. More generally, any instrument associated to a POVM $\{A_a\}$, i.e.\ any instrument with the property $\text{tr}[\mathcal I_a(\varrho)]=\text{tr}[\varrho A_a]$ holding for all states $\varrho$, is of the form $\mathcal I_a(\varrho)=\Lambda_a(\sqrt{A_a}\varrho\sqrt{A_a})$ where $\Lambda_a$ is an outcome-dependent quantum channel (from the input system to the output system) \cite{Pello4}. 
As an important special case, we highlight the von Neumann-L\"uders instrument $\mathcal I^{\text{vN-L}}_a(\varrho)=\sqrt{A_a}\varrho\sqrt{A_a}$, that is the most direct generalisation of the projection postulate, and can be seen as the least disturbing implementation of the POVM $\{A_a\}$, see Section~\ref{sec:retrieving}. Quantum instruments have been analysed intensively in the literature, see e.g. \cite{Pello4,Pello5,Pello6,Daviesbook,busch16,davies1970,ozawa,cycon,Holevo98}.

\subsection{Joint measurability}
\label{subsec:commNDJM}

There are three natural distinctions between classical and quantum properties of POVMs. These are given by non-commutativity, inherent measurement disturbance, and the impossibility of a simultaneous readout of the outcomes. Of these three, the last one has found the most profound role in quantum information theory, and consequently is our main focus. We start with the general notion of joint measurabilility and discuss the other two as special cases thereof.

The idea of joint measurability is to simulate the statistics of a set of measurements using only one measurement apparatus. This apparatus is described by a POVM $\{G_\lambda\}$ and its statistics in a state $\varrho$ read $p(\lambda|\varrho)=\text{tr}[G_\lambda \varrho]$. The set of measurements that we aim to simulate is described by a set of POVMs $\{A_{a|x}\}$. In this notation, $x$ labels the choice of the POVM and $a$ denotes the corresponding outcome. The simulation is done classically on the level of statistics and it is described by classical post-processings, i.e. conditional probabilities $\{p(a|x,\lambda)\}$. The simulation is successful if $\text{tr}[A_{a|x}\varrho]=\sum_\lambda p(a|x,\lambda)\text{tr}[G_\lambda \varrho]$ holds for any quantum state $\varrho$.

This leads to the formal definition of \textit{joint measurability}: A set of POVMs $\{A_{a|x}\}$ is said to be jointly measurable or compatible if there exists a POVM $\{G_\lambda\}$ and classical post-processings, i.e.\ a set of conditional probabilities, $\{p(a|x,\lambda)\}$ such that 
\begin{align}\label{eq:jmdefinition}
A_{a|x}=\sum_\lambda p(a|x,\lambda)G_\lambda.
\end{align}
In this case the POVM $\{G_\lambda\}$ is called a joint or parent measurement of the set $\{A_{a|x}\}$. Otherwise, the set $\{A_{a|x}\}$ is called not jointly measurable or incompatible.

\begin{figure}
\includegraphics[width=0.48\textwidth]{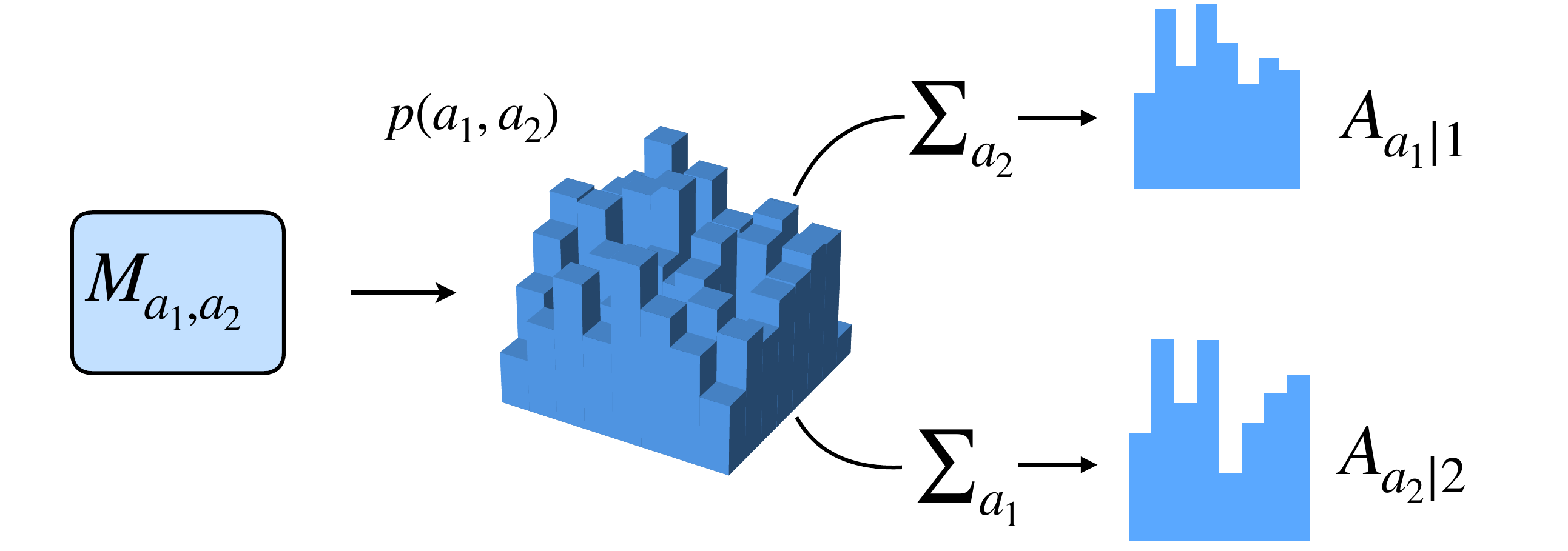}
\caption{Marginal form joint measurement of two POVMs. The data of the joint measurement is presented as a grid from which marginalisation gives the data of the original pair of measurements.}
    \label{fig:marginalJM}
\end{figure}

We note that the above definition is equivalent to the existence of a POVM $\{M_{\vec a}\}$, where $\vec a=(a_1,...,a_n)$ is a vector of the outcomes with the subindex referring to the measurement choice $x$, from which one gets the original POVMs as margins. More formally, one has 
\begin{align}\label{eq:MargJM}
A_{a|x}=\sum_{\vec a\in E_{a|x}} M_{\vec a},
\end{align}
where the set $E_{a|x}$ consists of those outcomes $\vec a$ of the joint measurement which include the outcome $a$ of the measurement $x$. As an example, in the case of two POVMs this reduces to $A_{a_1|1}=\sum_{a_2} M_{a_1,a_2}$ for all outcomes $a_1$ of the first measurement and $A_{a_2|2}=\sum_{a_1} M_{a_1,a_2}$ for all outcomes $a_2$ of the second measurement, see Fig.~\ref{fig:marginalJM}. In general, the measurement $\{M_{\vec{a}}\}$ can be viewed as a simultaneous readout of all its components in the sense that neglecting the data of all but one component gives the exact statistics of this component POVM. To see the equivalence between Eq.~(\ref{eq:jmdefinition}) and Eq.~(\ref{eq:MargJM}), one notes that clearly the marginal form is a special case of a general post-processing. For the other direction, one can set $M_{\vec a}:=\sum_\lambda[\Pi_x p(a_x|x,\lambda)]G_\lambda$, cf. \cite{ali09}.

The concept of joint measurability is easiest to illustrate with an example. Consider two measurements acting on a qubit system given by the POVM elements $A_{\pm|1}(\mu)=\frac{1}{2}\big(\openone\pm\mu\sigma_x\big)$ and $A_{\pm|2}(\mu)=\frac{1}{2}\big(\openone\pm\mu\sigma_z\big)$. These are noisy versions of the sharp spin measurements along the directions $x$ and $z$ with the parameter $1-\mu\in[0,1]$ describing the noise. An intuitive way to find a measurement with correct margins is to choose measurement directions that are in between $x$ and $z$, see also Section~\ref{sec:adaptive}, i.e. define a candidate joint measurement
\begin{align}
    M_{a_1,a_2}(\mu)=\frac14\Big(\openone+\mu(a_1\sigma_x+a_2\sigma_z)\Big).
\end{align}
This candidate has the correct margins and for $\mu\in[0,1/\sqrt 2]$ it is a POVM. Hence, the POVMs $A_{\pm|1}(\mu)$ and $A_{\pm|2}(\mu)$ are jointly measurable whenever $\mu\in[0,1/\sqrt 2]$. Moreover, whenever $\mu\in(0,1/\sqrt 2]$, the POVMs are non-commuting, but nevertheless a joint measurement exists. It can be shown that for $\mu>1/\sqrt 2$ the POVMs are not jointly measurable, see Section~\ref{sec-MUR}. This is a simple example of a joint measurement, but we note that in general their form can be complex and require an exponentially increasing number of outcomes \cite{Skrzypczyk2020}.

\subsection{Non-disturbance and commutativity}
\label{subsec:NDAC}

Joint measurability envelopes another central property of quantum measurements, that is the possibility of measuring POVMs in a sequence without disturbance. A POVM $\{A_{a_1|1}\}$ is said to be non-disturbing with respect to another POVM $\{A_{a_2|2}\}$ if there exists a sequential implementation in which neglecting the outcome of the first measurement $\{A_{a_1|1}\}$ does not affect the statistics of the subsequent measurement $\{A_{a_2|2}\}$. More precisely, one asks for the existence of an instrument $\{\mathcal I_{a_1}\}$ associated to $\{A_{a_1|1}\}$ such that $\sum_{a_1}\text{tr}[\mathcal I_{a_1}(\varrho)A_{a_2|2}]=\text{tr}[\varrho A_{a_2|2}]$ for all $\varrho$ and $a_2$. This notion generalises to more measurements straight-forwardly. It is clear that non-disturbance implies joint measurability by setting $\text{tr}[M_{a_1,a_2}\varrho]:=\text{tr}[\mathcal I_{a_1}(\varrho)A_{a_2|2}]$ for all $\varrho$. Interestingly, there are pairs of jointly measurable POVMs that do not allow for a non-disturbing sequential implementation \cite{heinosaari10}. However, jointly measurable pairs can always be measured in a sequence by performing a suitable instrument $\{\mathcal I_{a_1}\}$ of $\{A_{a_1|1}\}$ and a retrieving measurement $\{\tilde A_{a_2|2}\}$ after it, i.e.\ $\sum_{a_1}\text{tr}[\mathcal I_{a_1}(\varrho)\tilde A_{a_2|2}]=\text{tr}[\varrho A_{a_2|2}]$, see Section~\ref{sec:retrieving} and \cite{Pello7,heinosaari15b}. We note that in general the retrieving measurement is different from the original and it can be interpreted in two different ways. Either the measurement is a purely mathematical construction that relies on additional degrees of freedom on a larger Hilbert space, or it is a physical one, in which case one uses the output $a_1$ of the first measurement as an input for the second measurement. These cases are explained in more detail in Section~\ref{sec:retrieving}.

A historically relevant special case of joint measurability is that of commutativity. A set of POVMs $\{A_{a|x}\}$ is said to be commuting if $[A_{a|x},A_{b|y}]=0$ for all $a,b$ and $x\neq y$. It is clear that such a set allows a non-disturbing implementation by the use of the von Neumann-L\"uders instrument, and is jointly measurable with the product POVM $M_{\vec a}:=A_{a_1|1}\cdots A_{a_n|n}$, where $\vec a=(a_1,...,a_n)$. However, the inverse implications do not hold in general: As we have mentioned above, there are non-commuting POVMs that allow a joint measurement. Moreover, in \cite{heinosaari10} it was shown that when the Hilbert space dimension $d$ is equal to two, non-disturbance reduces to commutativity, but in systems with $d=3$ this is no longer true.

Although non-commutativity lacks an operational meaning in quantum measurement theory in general, with the exception of two-outcome (also called binary) measurements \cite{designolle21b}, it has been central for the development of quantum measurement theory. For example, non-disturbance and joint measurability are equivalent to commutativity in the case of PVMs (Hermitian operators). Also, a pair of POVMs is jointly measurable if and only if they have a common Naimark dilation in which the projective measurements on the dilation space commute, see Section~\ref{subsubsec:Naimark}.

\section{Characterizing joint measurability}
\label{sec-witness}

In this Section, we present the basic techniques for characterizing and quantifying 
incompatibility. First, we discuss analytical criteria for the qubit case and the 
connection to measurement uncertainty relations. Second, we explain the connections 
between joint measurability and the optimization method of semidefinite programming. 
This allows to introduce quantifiers of incompatibility. Third, we explain general 
methods to construct parent measurements and discuss connections with the Naimark 
extension of POVMs, which allows to formulate some results from a higher perspective.
Finally, we discuss algebraic characterizations of specific highly incompatible 
measurements in arbitrary dimensions. 

\subsection{Criteria for joint measurability and measurement uncertainty relations}
\label{sec-MUR}
Given the formal definition of joint measurability, one may ask for 
analytical criteria to determine whether two measurements are jointly 
measurable or not. In this subsection, we focus on analytical criteria
for measurements with two outcomes on a single qubit. As it turns out, there is an interesting connection to measurement
uncertainty relations.

For the case of qubits, the effects (i.e., positive operators bounded above by $\openone$) of all measurements may, up to 
normalization, be viewed as vectors on the Bloch sphere. So, for a 
two-outcome measurement $\{A_{\pm}\}$ we can write
\begin{equation}
A_{\pm} = \frac{1}{2} \big[(1 \pm \gamma) \openone \pm \vec{m} \cdot \vec{\sigma}\big]
\label{eq-measurement-bloch}
\end{equation}
with $\vec{m} \cdot \vec{\sigma} = m_x \sigma_x +m_y \sigma_y + m_z \sigma_z$.
Here, $\gamma$ is also called the bias of the measurement, while 
$\Vert \vec{m}\Vert$ is called the sharpness  \cite{busch16}. 

The first result on joint measurability of such measurements was 
already obtained by \citet{busch86}. He considered the case of two 
dichotomic measurements on a qubit, described by $\gamma_i$ and 
$\vec{m}_i$,  which are both unbiased ($\gamma_1 = \gamma_2 =0$). 
Then, he showed that these are jointly measurable, if and only if 
\begin{equation}
\Vert \vec{m}_1 + \vec{m}_2 \Vert + \Vert \vec{m}_1 - \vec{m}_2 \Vert \leq 2.
\label{eq-jm-busch}
\end{equation}
This relation has also been used to determine the probability of random measurements
to be incompatible \cite{zhang2019pre}.

For the case of two potentially biased measurements, this problem was 
considered by several authors independently at the same time \cite{0802.4248,0802.4167,yu10}. The resulting conditions are 
mathematically equivalent, but the most compact form was derived by 
\citet{yu10}. For that, one defines the auxiliary quantities
\begin{equation}
F_i = \frac{1}{2}
\big[
\sqrt{(1+\gamma_i)^2- \Vert \vec{m}_i\Vert^2}
+\sqrt{(1-\gamma_i)^2- \Vert \vec{m}_i\Vert^2}
\big]
\end{equation}
for $i=1,2$. Then, the measurements $\{A_{\pm|1}\}$ and $\{A_{\pm|2}\}$ 
are jointly measurable if and only if 
\begin{equation}
(1-F_1^2 - F_2^2)\big(1- \frac{\gamma_1^2}{F_1^2} - \frac{\gamma_2^2}{F_2^2}\big)
\leq (\vec{m}_1 \cdot \vec{m}_2 - \gamma_1 \gamma_2 )^2.
\end{equation}
Finally, several works extended the condition in Eq.~(\ref{eq-jm-busch}) 
to three unbiased measurements. This has first been done for unbiased 
measurements in orthogonal directions \cite{busch86, brougham2007} and for three 
measurement directions in angles of $2\pi/3$ in a plane \cite{liang2011}. 
For three general measurements a necessary condition was found in 
\cite{pal2011}, which was then shown to be sufficient for unbiased measurements 
by \citet{yu13}. It reads as follows: For a set of vectors $\{\vec{v}_k\}$ one defines the 
Fermat-Toricelli vector $\vec{v}_{FT}$ as the vector minimizing the sum 
of the distances $\sum_k \Vert \vec{v}- \vec{v}_k \Vert$. Then, three 
unbiased measurements on a qubit  are jointly measurable, if and only if 
\begin{equation}
\sum_{k=0}^3 \Vert \vec{T}_k - \vec{T}_{FT} \Vert \leq 4,
\end{equation}
where $\vec{T}_{FT}$ is the Fermat-Toricelli vector of the four vectors
$\vec{T}_0 = \vec{m}_1+\vec{m}_2+\vec{m}_3$ and $\vec{T}_k= 2\vec{m}_k-\vec{T}_0$
for $k=1,2,3.$

So far, we discussed criteria of joint measurability for pairs or triples
of measurements. This leads to the question, which joint measurability structures in a set of POVMs are possible. For instance, one may ask for
a triple of measurements, where each pair is jointly measurable, but all
three are not jointly measurable. Such an example has been constructed
by \citet{heinosaarifop2008}. In fact, for large sets of measurements, 
arbitrary joint measurability structures can be realized \cite{kunwal2014, andrejic2020}.

The previous exact solutions of the joint-measurability problem for certain 
instances are not only of mathematical beauty, but they are also relevant 
for deriving measurement uncertainty relations. We have seen in Eq.~(\ref{eq-robertson}) that the commutator of two projective measurements occurs naturally 
in the formulation of the Robertson uncertainty relation. The Robertson 
uncertainty relation is a preparation uncertainty relation in the sense 
that it constrains the ability to prepare states which are close to common 
eigenstates of the observables, but this is not directly related to the 
measurement process of the observables.

\begin{figure}
\includegraphics[width=0.48\textwidth]{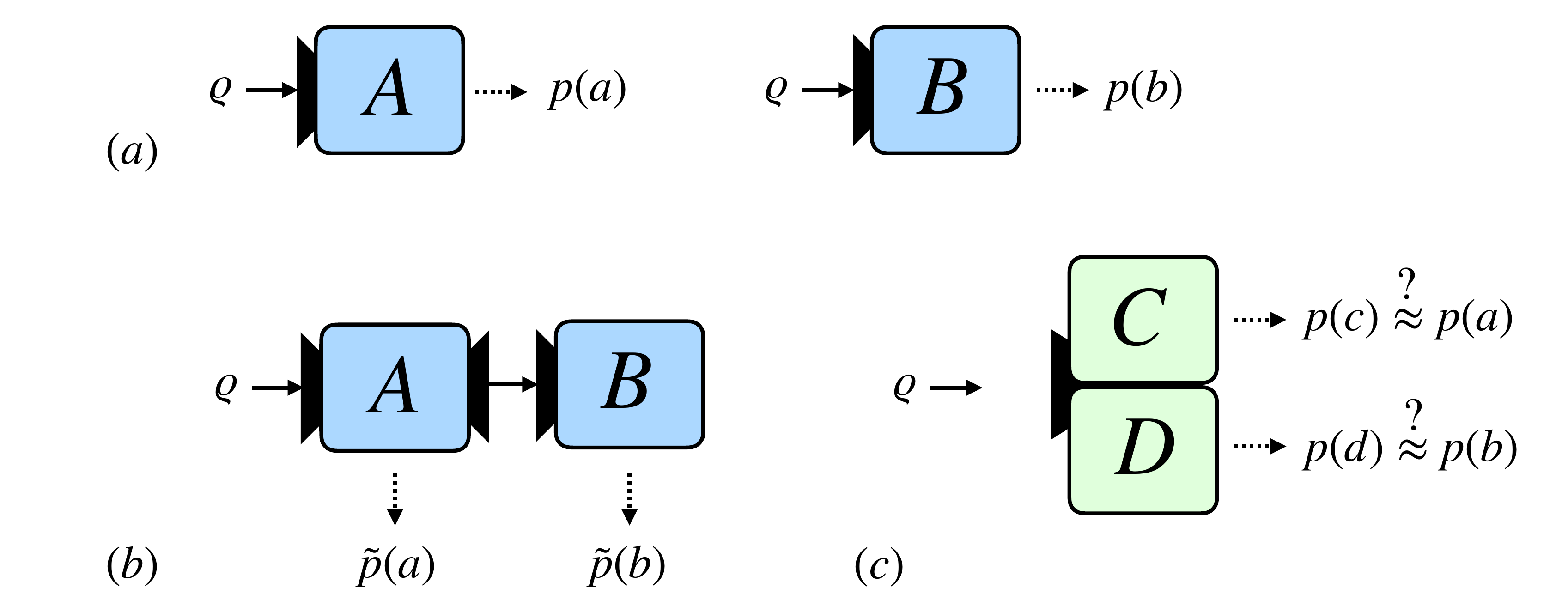}
\caption{Schematic view of measurement uncertainty relations. 
(a) For any quantum state $\rho$ the measurement $A$ yields results 
with a probability distribution $p(a)$, similarly $B$ delivers a 
distribution $p(b)$. (b) A possible way to study the disturbance
of $B$ through $A$ is to measure in sequence and compare the
distributions $\tilde p (a)$ and $\tilde p (b)$ with $p(a)$ and 
$p(b)$. Also the reverse order of measurements can be considered.
(c) Since the measurements in (b) give other probability distributions
than in (a), they can be described by different POVMs, $C$ and $D$. But
then, $C$ and $D$ are jointly measurable by construction. The question
arises: What is the best approximation of $A$ and $B$ by two jointly
measurable POVMs $C$ and $D$?}
\label{fig:measurement-ur}
\end{figure}

In recent years, the notion of measurement uncertainty relations has been 
used to quantify potential constraints and disturbances during the measurement 
process \cite{werner04,buschrmp2014}. We can use the above condition (\ref{eq-jm-busch}) to 
explain this concept in a simple setting \cite{bullock2018}. Assume that two projective measurements $A$ and $B$ on a qubit shall be implemented simultaneously. Since 
they may not be jointly measurable, one has to implement two POVMs $C$ and $D$ 
as approximations of $A$ and $B$, respectively, where $C$ and $D$ are jointly measurable, see also Fig.~\ref{fig:measurement-ur}. This, of course, introduces 
an error, which may be quantified by the difference of the probabilities of one 
result $D^2(A,C) = 4 |p(A=+)-p(C=+)|$, which is a simple case of the so-called Wasserstein distance between two probability distributions. If one considers the 
worst case, and maximizes this error over all quantum states, one finds the relation
\begin{equation}
D^2(A,C) + D^2(B,D) \geq \sqrt{2} \big( 
\Vert \vec{a} + \vec{b} \Vert + \Vert \vec{a} - \vec{b} \Vert - 2
\big),
\end{equation}
where $\vec{a}$ and $\vec{b}$ are the Bloch vectors of the measurements, such 
as $\vec{m}$ in Eq.~(\ref{eq-measurement-bloch}). This shows that 
Eq.~(\ref{eq-jm-busch}) can result in uncertainty relations similar as the 
commutator in Eq.~(\ref{eq-robertson}).

\subsection{Quantification of incompatibility}
\label{sec-SDP}
\subsubsection{Joint measurability as a semidefinite program}

Given a set of measurements, the existence of a joint measurement 
can be decided through convex optimization techniques that we will 
explain in the following. It is instructive to consider the case 
of 
two arbitrary two-outcome (or binary) POVMs first \cite{wolf09}. Let 
$A_1 = \{A_{+|1}, A_{-|1}\}$ and $A_2 = \{A_{+|2}, A_{-|2}\}$ be two 
POVMs in dimension $d$. They are jointly measurable if there 
exists a four-outcome measurement $M_{a_1,a_2}$ with $a_1,a_2=\pm$ 
such that $M_{++}+M_{+-}=A_{+|1}$ and $M_{++}+M_{-+}=A_{+|2}$. By 
writing the elements of the parent POVM in terms of $A_1$ and $A_2$, 
one realizes that their compatibility is equivalent to the existence 
of a positive semidefinite operator $M_{++}$ for which 
\begin{equation}
A_{+|1}+A_{+|2}-\openone \leq M_{++} \leq A_{+|x}
\label{eq-simple-sdp}
\end{equation}
for $x=1,2$. The problem of deciding if such an operator exists 
can be cast as a semidefinite program (SDP) by minimizing a real 
number $\gamma$ subject to the constraints that 
$A_{+|1}+A_{+|2} \leq \gamma \openone +M_{++}$ 
and $0\leq M_{++}\leq A_{+|x}$ for $x=1,2$. If $\gamma =1$ 
can be reached, then the measurements $A_1$ and $A_2$ are 
jointly measurable.

In its most general form, an SDP can be written as
\begin{align}
\max  \quad & \tr[AX] \\
\text {s.t.} \quad & \Phi(X)=B, \\
& X\geq 0,
\end{align}
where $A$ and $B$ are Hermitian operators and $\Phi$ is a Hermiticity-preserving 
linear map. Note that in the literature SDPs are also frequently written as a 
maximization of the function $\sum_i c_i x_i$ over real variables $x_i$, subject 
to the constraint that $F_0 +\sum_i x_i F_i \geq 0$ is a positive semidefinite matrix, 
and the $F_i$ are hermitian matrices.

The theory of convex optimization and in particular of semidefinite programming 
is very well developed \cite{boyd04, GaertnerMatousek2012} and is a frequently 
used tool in quantum information theory \cite{Watrous2018}. In fact, the optimization 
problem above can be easily solved using available software such as CVX \cite{cvx} 
or MOSEK \cite{mosek}. SDPs enjoy many properties that make them useful as 
a mathematical tool. For instance, each SDP can be associated to its so-called 
\emph{dual program}, which reads
\begin{align}
\min \quad & \tr[BY] \\
\text {s.t.} \quad & \Phi^\dagger(Y)\geq A, \\
& Y=Y^\dagger,
\end{align}
where $\Phi^\dagger$ denotes the adjoint map of $\Phi$ defined by 
$\tr[T\Phi(X)]=\tr[\Phi^\dagger(T)X]$. An important property of many 
SDPs is known as \emph{strong duality}, which refers to the fact that 
under certain conditions known as \emph{Slater's conditions} the optimal 
values of the primal and the dual problem coincide.

Concerning general POVMs, one can formulate the SDP based on the following
considerations. First, one observes that the classical postprocessing in
Eq.~(\ref{eq:jmdefinition}) can be chosen deterministic, i.e., 
$p(a|x,\lambda) = D(a|x,\lambda)$ takes only values zero and one~\cite{ali09}. Then, 
there is only a finite number of such postprocessings. So, for a set $\{A_{a|x}\}$ 
of POVMs the following SDP decides whether or not they are jointly measurable:
\begin{align}
\label{eq:JMSDP}
\text{given} \quad &\{A_{a|x}\}_{a, x}, \{D(a|x,\lambda)\}_\lambda \\
\max_{\{G_{\lambda}\}}\quad & \mu \\
\text {s.t.}\quad & \sum_{\lambda} D(a|x,\lambda) G_{\lambda}=A_{a|x} \quad \forall a, x \\
& G_{\lambda} \geq \mu \openone \quad  \sum_\lambda G_\lambda=\openone.
\end{align}
This optimization is performed for each fixed deterministic post-processing $\{D(a|x,\lambda)\}_\lambda$. 
Clearly, if this optimization results in a value of $\mu$ strictly less than zero the positivity 
constraint on the joint observable cannot be fulfilled, which proves incompatibility. 
Otherwise, a joint observable is found which proves joint measurability.

\subsubsection{Various quantifiers of incompatibility}

Typically one is not only interested in answering the question whether or 
not a set of measurements is incompatible, but one is also interested 
in quantifying to what extent the measurements are incompatible. Similarly, 
in entanglement theory one not only asks if a state $\varrho$ is entangled 
or separable, but also one is interested in how close a state is to being separable. 
This can be done by adding a certain amount of noise until an entangled state becomes 
separable. Different types of noise lead to different quantifiers including the 
\emph{(generalized) entanglement robustness} \cite{Vidal1999,Steiner2003,Brandao2005,guehne09}
or the \emph{best separable approximation} 
introduced by~\citet{LewensteinSanpera98}. 

Similarly, for measurement incompatibility one can ask how \emph{close} a set 
of POVMs is to the set of compatible sets of POVMs by means of adding a 
certain type of noise. Consider a set of incompatible POVMs to which we add 
an amount $p\in \left[0,1\right]$ of classical noise. The resulting POVMs are 
thus given by $A_{a|x}^{(p)}=(1-p )A_{a|x}+p {\openone}/{\abs{a}}$, where $\abs{a}$ denotes the number of outcomes, and the quantity
\begin{equation}
R_{\rm inc}^{\rm n} (A_{a|x})=\inf\big\{p\,\big\vert\, A_{a|x}^{(p)}\,\text{ are compatible}\big\}
\label{eq-random-robustness}
\end{equation}
is called the incompatibility~\emph{noise robustness}, cf.~\citet{Heinosaari2015b}, which is one minus the~\emph{incompatibility random robustness} in ~\citet{designolle19b}. It was shown by \citet{Heinosaari2015b} that 
this quantity is an \emph{incompatibility monotone}, since it fulfils the following properties: 
(i) it vanishes on compatible sets, (ii) it is symmetric under exchange of measurements, and 
(iii) it does not increase under pre-processing by a quantum channel.

\begin{figure}[t]
\includegraphics[width=0.8\columnwidth]{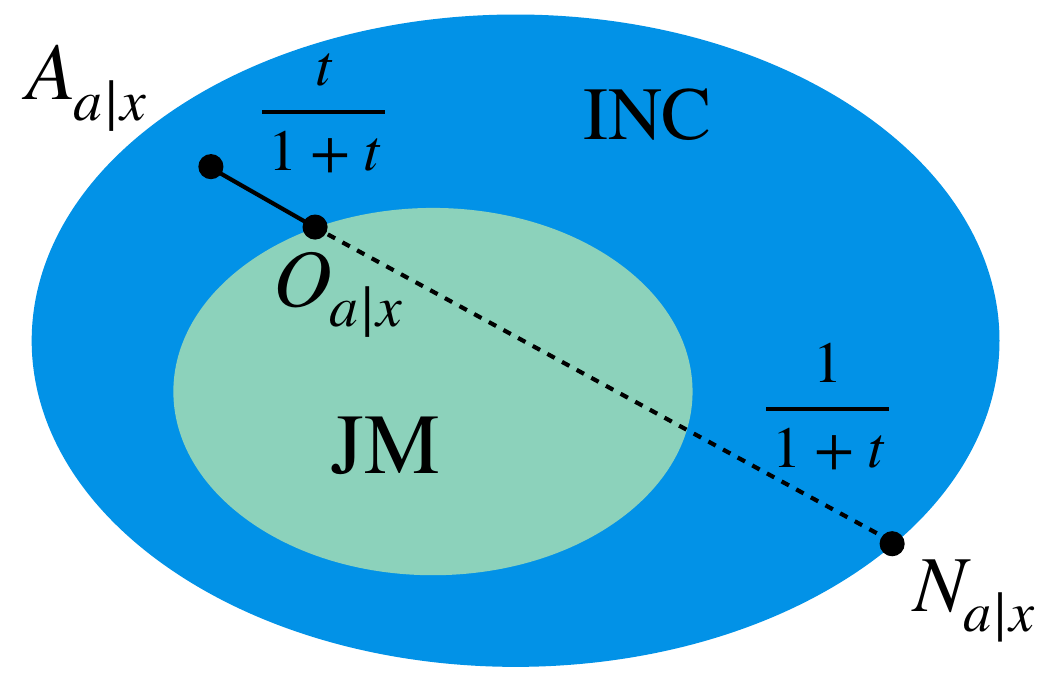}
\caption{Illustration of the incompatibility robustness. JM denotes compatible sets and INC denotes incompatible sets. The POVMs $\{A_{a|x}\}$ are mixed with arbitrary "noise" POVMs $\{N_{a|x}\}$ such that their mixture is compatible. The incompatibility robustness is the smallest $t$ that realizes such a decomposition, cf. Eq.~\eqref{eq:robustness}.
\label{fig-robustness}}
\end{figure}

When other types of noise are considered one arrives at similar quantities that 
all share similar properties, e.g., they act as monotones under certain transformations. One example is the \emph{incompatibility weight} \cite{pusey15}, 
which is analogous to the \emph{steering weight} defined by~\citet{gallego15}. The 
incompatibility weight of a set $\{A_{a|x}\}$ of POVMs is the smallest value of $\nu$ 
for which the decomposition $A_{a|x}=\nu N_{a|x}+(1-\nu)O_{a|x}$ exists, 
where $N_{a|x}$ is an arbitrary "noise" POVM, and $\{O_{a|x}\}$ are jointly measurable. 
More precisely,
\begin{equation}\label{eq:weight}
W_{\rm inc}(A_{a|x})=\inf\big\{\nu\geq 0\, \vert\, \frac{A_{a|x}-\nu N_{a|x}}{1-\nu}=O_{a|x}\big\}.
\end{equation}
In entanglement theory this quantity is known as the best separable approximation 
\cite{LewensteinSanpera98}. It was furthermore shown that the incompatibility 
weight is a monotone when more general transformations than quantum pre-processing 
are allowed \cite{pusey15}. More precisely, the incompatibility weight is a monotone 
under compatibility non-decreasing operations (CNDO) that consist of pre-processing 
by a quantum instrument and conditional classical post-processing.

A similar construction is the 
\emph{incompatibility robustness}~\cite{haapasalo15a, uola15},
defined by
\begin{equation}
\label{eq:robustness}
R_{\rm inc}(A_{a|x})=\inf\{t\geq 0\, \vert\, \frac{A_{a|x}+t N_{a|x}}{1+t}=O_{a|x}\},
\end{equation}
where $\{N_{a|x}\}$ is any set of POVMs and $\{O_{a|x}\}$ are jointly measurable, 
see also Fig.~\ref{fig-robustness}. Similar to the incompatibility weight, the incompatibility robustness is a monotone under CNDO.

These quantifiers are all based on the convex distance of incompatible POVMs 
to the set of jointly measurable ones under the addition of different types 
of noise. In particular, all these distances can be evaluated numerically 
as they fall in the framework of SDPs explained above. For instance, it was 
shown, following the construction of the steering robustness \cite{piani15b}, 
that the incompatibility robustness can be cast as the following 
optimization problem \cite{uola15}
\begin{align}\label{eq:JMRobustness}
\text{min} \quad &\frac{1}{d} \sum_{\lambda} \operatorname{tr}\left[G_{\lambda}\right] \\
\text{s.t.} \quad &\sum_{\lambda} D(a | x, \lambda) G_{\lambda} \geq A_{a \mid x} \quad \forall a, x, \\
&G_{\lambda} \geq 0 \\
&\sum_{\lambda} G_{\lambda}=\frac{\openone}{d}\Big(\sum_{\lambda} \operatorname{tr}\left[G_{\lambda}\right]\Big).
\end{align}
A recent and more detailed review on the numerical evaluation of robustness based 
incompatibility measures can be found in Ref.~\cite{cavalcanti17}.

\subsection{Constructing joint measurements}
\label{sec-constructing-JM}
\subsubsection{Adaptive strategy}
\label{sec:adaptive}
An intuitive way to build joint measurements for given POVMs was presented in \cite{uola16}. The idea is to exploit 
classical randomness between measurements that are in some sense similar 
to the original ones. In the simplest case of two measurements $\{A_{a|x}\}$ with 
$x=1,2$, one can flip a coin to decide which measurement to perform, and assign 
an outcome to the other measurement based on the gained information, i.e. 
classically post-process the outcome. However, in many scenarios it is better 
to flip a coin between some other measurements than the original pair.

\begin{figure}[t]
    \includegraphics[width=0.48\textwidth]{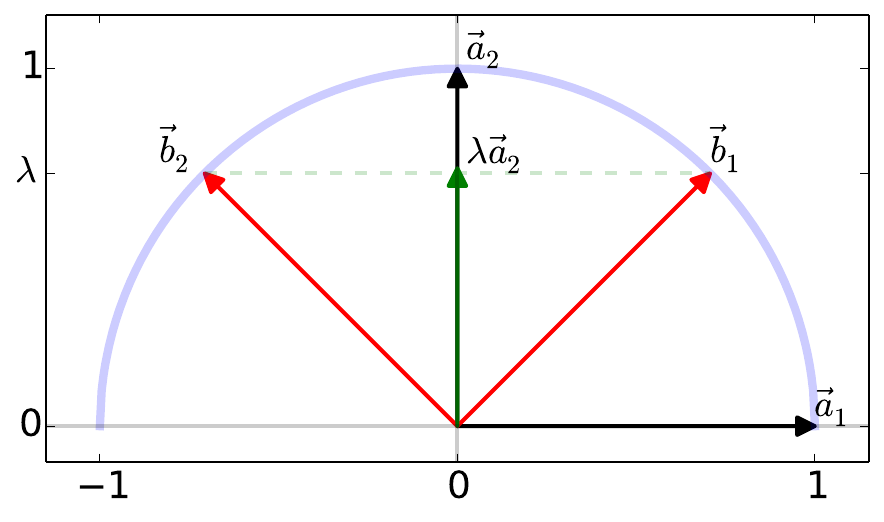}
    \caption{An equally weighted coin toss between the auxiliary elements $B_{+|1}(\lambda,\lambda)$ and $B_{+|2}(\lambda,\lambda)$ with $\lambda=1/\sqrt{2}$ results in an effect with the Bloch vector $\lambda\vec a_2=\lambda(0,0,1)=(0,0,1/\sqrt 2)$. Here $\vec b_1$ and $\vec b_2$ are the Bloch vectors of $B_{+|1}(\lambda,\lambda)$ and $B_{+|2}(\lambda,\lambda)$. The vectors $\vec a_1$ and $\vec a_2$ represent the directions of the original measurements.}
    \label{fig:adaptive}
\end{figure}

To illustrate the technique, we define a pair of noisy spin measurements 
as $A_{\pm|1}(\mu)=\frac12\big(\openone\pm\mu\sigma_x\big)$ and 
$A_{\pm|2}(\lambda)=\frac12\big(\openone\pm\lambda\sigma_z\big)$. The 
auxiliary measurements are defined as 
$B_{\pm|1}(\mu,\lambda)=\frac12\big[\openone\pm(\mu\sigma_x+\lambda\sigma_z)/N\big]$ 
and $B_{\pm|2}(\mu,\lambda)=\frac12\big[\openone\pm(\mu\sigma_x-\lambda\sigma_z)/N\big]$, 
see also Fig.~\ref{fig:adaptive} for the case $\mu=\lambda$. Here $N=\sqrt{\mu^2+\lambda^2}$ 
is the norm of the Bloch vector, which guarantees the positivity of the effects of the auxiliary measurements.
Flipping a coin between these measurement and making the obvious assignments of values leads 
to a marginal form joint POVM with the effects
\begin{align}
    M_{++}(\mu,\lambda)&=\frac12 B_{+|1}(\mu,\lambda),\quad M_{--}(\mu,\lambda)=\frac12 B_{-|1}(\mu,\lambda)\\
    M_{-+}(\mu,\lambda)&=\frac12 B_{-|2}(\mu,\lambda),\quad M_{+-}(\mu,\lambda)=\frac12 B_{+|2}(\mu,\lambda).
\end{align}
More compactly, we have $M_{i,j}=\frac14\big(\openone+(i\mu\sigma_x+j\lambda\sigma_z)/N\big)$. Clearly this joint measurement gives the noisy versions $A_{\pm|1}(\mu/N)$ and $A_{\pm|2}(\lambda/N)$ as marginals. For example, in the case $\mu=\lambda=1/\sqrt 2$ we get $N=1$, which corresponds to the optimal threshold in Eq.~(\ref{eq-jm-busch}), cf. Fig~\ref{fig:adaptive}.

The above approach for finding joint measurements is entitled an adaptive strategy \cite{uola16}, as it uses the gained information to assign values to other measurements. In principle, one can use unbiased coins and a different numbers of auxiliary measurements, and every such scenario will lead to a joint measurement of some POVMs. Whereas random or uneducated guesses for the auxiliary measurements are not guaranteed to give a good joint measurement, optimal auxiliary measurements are rather straight-forward to find in scenarios with symmetry, such as in the case of pairs of mutually unbiased bases (MUBs for short) \cite{uola16}, see also Section \ref{sec:HDsymm} for explicit results on symmetric measurement sets.

\subsubsection{Operator measure with correct marginals}
\label{sec:wmeasure}

\citet{jae19} proposed to build joint measurements for pairs of 
measurements using a specific ansatz. For a pair of POVMs $\{A_a\}$ and $\{B_b\}$ 
both with $n$ values, the authors define a so-called $W$-measure as
\begin{align}
\label{eq-jaeiteration}
W_{ab}=C_{ab}+\frac{1}{n}(A_a-\sum_j C_{aj})+\frac{1}{n}(B_b-\sum_i C_{ib}),
\end{align}
where $\{C_{ab}\}$ is an arbitrary POVM. Clearly, $\{W_{ab}\}$ has the original 
pair of POVMs as its marginals, but its elements are not required to be positive 
semi-definite. Still, this ansatz allows already a numerical treatment. If
one takes $\{C_{ab}\}$ to be the parent POVM $\{M_{ab}\}$, then also $\{W_{ab}\} 
= \{M_{ab}\}$. So, Eq.~(\ref{eq-jaeiteration}) 
can be used to formulate an iteration, with the desired parent POVM as one fixed point.

Moreover, it is straight-forward to show that instead of using 
the POVM $\{C_{ab}\}$ one can parametrise $W$-measures as
\begin{align}
W_{ab}=\frac{1}{n}(A_a+B_b)-\Omega_{ab},
\end{align}
where $\{\Omega_{ab}\}$ are Hermitian operators with the property 
$\sum_a\Omega_{ab}=\sum_b\Omega_{ab}=\openone/n$. The authors note 
that the original pair is jointly measurable if and only if there 
exists a collection $\{\Omega_{ab}\}$ such that the corresponding 
collection $\{W_{ab}\}$ is positive semi-definite.

This allows to introduce the negativity of a $W$-measure as 
$\mathcal N:=\frac{1}{d}\sum_{a,b,k}(|\lambda_{ab}^k|-\lambda_{ab}^k)$, 
where $\{\lambda_{ab}^k\}_k$ are the eigenvalues of $W_{ab}$. Clearly, 
$\mathcal N =0$ for some $\{\Omega_{ab}\}$ if and only if the POVMs
are jointly measureable. Note that this enables to decide joint measurability
by a direct minimization, which may be solved analytically for special cases 
without using SDPs.

Indeed, \citet{jae19} minimise $\mathcal N$ over all collections $\{\Omega_{ab}\}$ 
for two important cases, namely general unbiased qubit POVMs with two outcomes and special qubit so-called trinary POVMs with three outcomes. The former one results in the known Busch criterion in 
Eq.~(\ref{eq-jm-busch}), and the latter results in
\begin{align}
\mathcal N_{\text{min}}=\text{max} (\frac{1}{9}\sum_{a,b}\|\vec m_a+\vec n_b-\vec\theta_{ab}\|-1,0)
\end{align}
for two trinary POVMs with the Bloch vectors $\{\vec m_a\}$ of the POVM 
$\{A_a\}$ and $\{\vec n_b\}$ of the POVM $\{B_b\}$, where the three effects
for each of the measurements all lie in the same plane of the Bloch sphere. 
Here $\vec{\theta}_{ab}=\vec m_{2(a+b)}+\vec n_{2(a+b)}$. This can be translated 
into a condition on joint measurability, reading
\begin{align}
\sum_{a,b}\|\vec m_a+\vec n_b-\vec \theta_{ab}\|\leq 9.
\end{align}

\subsubsection{Naimark strategy}
\label{subsubsec:Naimark}
Here we review an analytical technique for characterising all 
possible parent POVMs related to a given measurement \cite{pellonpaa14,Pello7}. 
This relies on extending the POVMs to PVMs on a larger system by the 
so-called Naimark extension. The method shows a tight connection between 
joint measurability and commutativity in the extended Hilbert space 
picture. As a direct consequence, the technique gives an important structural 
result: All POVMs jointly measurable with a given rank-1 POVM are its 
post-processings.

Let us first introduce the desired Naimark extension \cite{naimark, peresbook}. Let $\{A_a\}$ be a POVM with $n$ outcomes in a $d$-dimensional 
system described by the Hilbert space $\mathcal{H}$. We can 
dilate $\{A_a\}$ to a larger Hilbert space as follows:
Each effect of the POVM can be written as
$A_a=\sum_{k=1}^{m_a}|d_{ak}\rangle\langle d_{ak}|$, 
where the vectors $|d_{ak}\rangle=\sqrt{\lambda_{ak}}|\varphi_{ak}\rangle$ for 
$k=1,\dots,m_a$ are the (unnormalized) eigenvectors associated with the nonzero 
eigenvalues $\lambda_{ak}$ and $m_a$ is the rank of $A_a$.
Consequently, $\{\ket{\varphi_{ak}}\}_{k=1}^{d}$ is an orthonormal 
eigenbasis of $A_a$. 

In order to write down the extension, 
let $\hil_\oplus$ be a $(\sum_{a}m_a)$-dimensional Hilbert space 
with an orthonormal basis $\{|e_{ak}\rangle\}_{a,k}$ and consider 
the PVM $P_a=\sum_{k=1}^{m_a}|e_{ak}\rangle\langle e_{ak}|$ with 
$n$ outcomes $a$. Then we can define the map
$J=\sum_{a=1}^n\sum_{k=1}^{m_a}|e_{ak}\rangle\langle d_{ak}|$,  
which is an isometry, since $J^\dagger J=\openone$ on $\mathcal{H}$.
Now $A_a=J^\dagger P_a J$ so that the triplet $(\hil_\oplus,J,\{P_a\})$ 
is a Naimark dilation of $\{A_a\}$. This dilation is minimal, meaning
that the set of vectors $\big\{P_aJ\ket{\psi}\big|a,\ket{\psi}\big\}$ spans $\hil_\oplus$ \cite{pellonpaa14,Pello7}. Note that $\{A_a\}$ is sharp exactly when 
$J$ is unitary. In this case, we may identify $\{A_a\}$ with $\{P_a\}$.

Based on this Naimark dilation, one can obtain several structural insights.
If $\{B_b\}$ is a POVM jointly measurable with $\{A_a\}$ then any joint 
measurement $\{M_{ab}\}$ is of the form $M_{ab}=J^\dagger P_a\tilde B_b J$,
where $\{\tilde B_b\}$ is a unique POVM of $\hil_\oplus$ which commutes with 
$P_a$, i.e.\ $P_a\tilde B_b=\tilde B_b P_a$ \cite{Pello7,pellonpaa14}. 
The uniqueness follows since the dilation is minimal. This gives an effective 
method for constructing all POVMs jointly measurable with $\{A_a\}$: They 
are of the form $B_b=J^\dagger\tilde B_b J$. 

We have two immediate special cases: If $\{A_a\}$ is sharp (i.e.\, $J$ 
is unitary) then $M_{ab}=J^\dagger P_a J J^\dagger\tilde B_b J=A_a B_b=B_b A_a$, 
so that the POVMs must commute. If $\{A_a\}$ is of rank-1, i.e.\ any $m_a=1$
and $A_a=|d_{a}\rangle\langle d_{a}|$ and $P_a=|e_{a}\rangle\langle e_{a}|$ \cite{Pello9}, 
then each $\tilde B_b$ is diagonal in the basis $\{|e_{a}\rangle\}$, so one
can write $\tilde B_b=\sum_a p(b|a)|e_{a}\rangle\langle e_{a}|$. From this one 
obtains $M_{ab}=p(b|a)A_a$ where $p(b|a)$ is a conditional probability. 
Thus, any POVM $\{B_b\}$ that is jointly measurable with $\{A_a\}$
is a classical postprocessing, $B_b=\sum_a p(b|a)A_a$.
Finally, we note that, if $M_{ab}=J^\dagger P_{ab} J$ is a Naimark dilation of a joint POVM of $\{A_a\}$ and $\{B_b\}$ then we get a common Naimark dilation for $A_a=J^\dagger (\sum_b P_{ab})J$ and 
$B_b=J^\dagger (\sum_a P_{ab})J$ where the marginal PVMs commute.

\subsection{High-dimensional measurements and symmetry}\label{sec:HDsymm}
So far, we have presented methods for the characterization of 
joint measurability which are mainly applicable to low-dimensional
systems. The explicit criteria in Section \ref{sec-MUR} were formulated for 
qubits and the computational approaches in Sections \ref{sec-SDP} and \ref{sec:wmeasure} 
are naturally restricted due to numerical limitations. As we explain now, 
for high-dimensional systems often symmetries and algebraic relations
can be used to characterize joint measurability.

To start, let us consider the case of measurements in mutually 
unbiased bases (MUBs). Two bases $\ket{\psi_i}$ and $\ket{\phi_j}$ 
of a $d$-dimensional space are called mutually unbiased, if they obey
\begin{equation}
|\braket{\psi_i}{\phi_j}|^2 = \frac{1}{d} 
\end{equation}
for all $i,j$. To give an example, the eigenstates of the three Pauli
matrices form a triple of MUBs. In fact, MUBs can be seen as a 
generalization of the Pauli matrices to higher dimensions, and as 
such one may expect MUBs to correspond to highly incompatible 
measurements. Independent of that, MUBs are relevant for various
quantum information processing tasks like quantum tomography or
quantum key distribution. It is known that for a given 
$d$ maximally $d+1$ MUBs can exist, but whether this bound can 
be reached is an open problem and, indeed, one of the hard problems 
in quantum information theory, see also \cite{bengtsson2007, horodecki2020open} 
for an overview.

In order to quantify the incompatibility of general MUBs, 
\citet{designolle19a} proceeded as follows. First, as a quantifier
they used a variant of incompatibility noise robustness
as introduced in Eq.~(\ref{eq-random-robustness}), by considering
the noisy POVM $A^\eta_{a|x} = \eta A_{a|x} + (1-\eta) \tr(A_{a|x}) \openone/d$
and asking for the maximal $\eta^*$, such that the POVMs $A^\eta_{a|x}$ 
are compatible. Such robustness is also called the depolarising or the white noise robustness. The optimal $\eta^*$ can be computed by an SDP. For deriving
upper bounds on $\eta^*$, one can consider the dual optimization 
problem, which is a minimization problem. Inserting a specific instance
of the dual variables results in an analytical upper bound on $\eta^*.$ 
For instance, for $k$ projective measurements where the projectors are
of rank one, one finds that 
\begin{equation}
\eta^* 
\leq 
\eta_{\rm up}
= 
\frac{\lambda-k/d}{k-k/d},
\label{eq-mub-estimates}
\end{equation}
where $\lambda$ is the largest eigenvalue of an operator $X$ 
that can be obtained by selecting one outcome per measurement,
that is, $X = \sum_{x=1}^k A_{j_x|x}$ for some $\vec{j}.$ 
Of course, also a set of $k$ MUBs can be viewed as measurements, and 
for prime power dimensions, there is an explicit construction of
$d+1$ MUBs due to \citet{wootters1989}. It turns out that when taking
$k=2$, $k=d$, or $k=d+1$ of these MUBs one finds that $\eta^* = \eta_{\rm up},$
so in this case these MUBs are maximally incompatible. It is important
to stress, however, that in general two sets of MUBs can be inequivalent
(in the sense that they are not connected by a unitary transformation
or permutation), and in general MUBs do not reach the bound in 
Eq.~(\ref{eq-mub-estimates}). Still, one can prove for general MUBs 
a lower bound on $\eta^*$ \cite{designolle19a}, as well as upper and 
lower bounds for general measurements \cite{designolle19b}

This result begs the question, which measurements are most incompatible
for a given quantifier of measurement incompatibility. This has been
studied in detail in \cite{designolle19b, bavaresco17}. Not surprisingly, the most 
incompatible pair of measurements depends on the chosen quantifier 
and sometimes other measurements beside MUBs are the most incompatible 
ones. 

Finally, one may ask how interesting sets of measurements, e.g., with 
a high incompatibility or other nice properties can be identified 
from abstract principles. This problem has been addressed by 
\citet{nguyen20}. There, sets of measurements has been studied 
from group theoretic perspectives. In fact, starting from a 
complex reflection group $G$ and with a given representation one
can construct a measurement assemblage (i.e., a set of measurements) with certain symmetries. 
These assemblages have then often interesting physical properties, 
e.g., a high incompatibility.

\section{Incompatibility and quantum information processing}
\label{sec-correlations}

Measurement incompatibility is inherently linked to the non-classical 
character of quantum correlations. Indeed, for many scenarios it is 
easy to see that incompatibility of the performed measurements is 
necessary for displaying nonclassical correlations. Since 
such correlations are required for tasks like quantum key distribution 
or quantum metrology, these connections highlight the resource aspect of 
measurement incompatibility. In this section, we describe different 
forms of non-classical correlations as well as other phenomena, 
for which the incompatibility of measurements is essential.

	\subsection{Bell nonlocality}
	\label{sec-non-locality}
\begin{figure}
\includegraphics[width=0.45\textwidth]{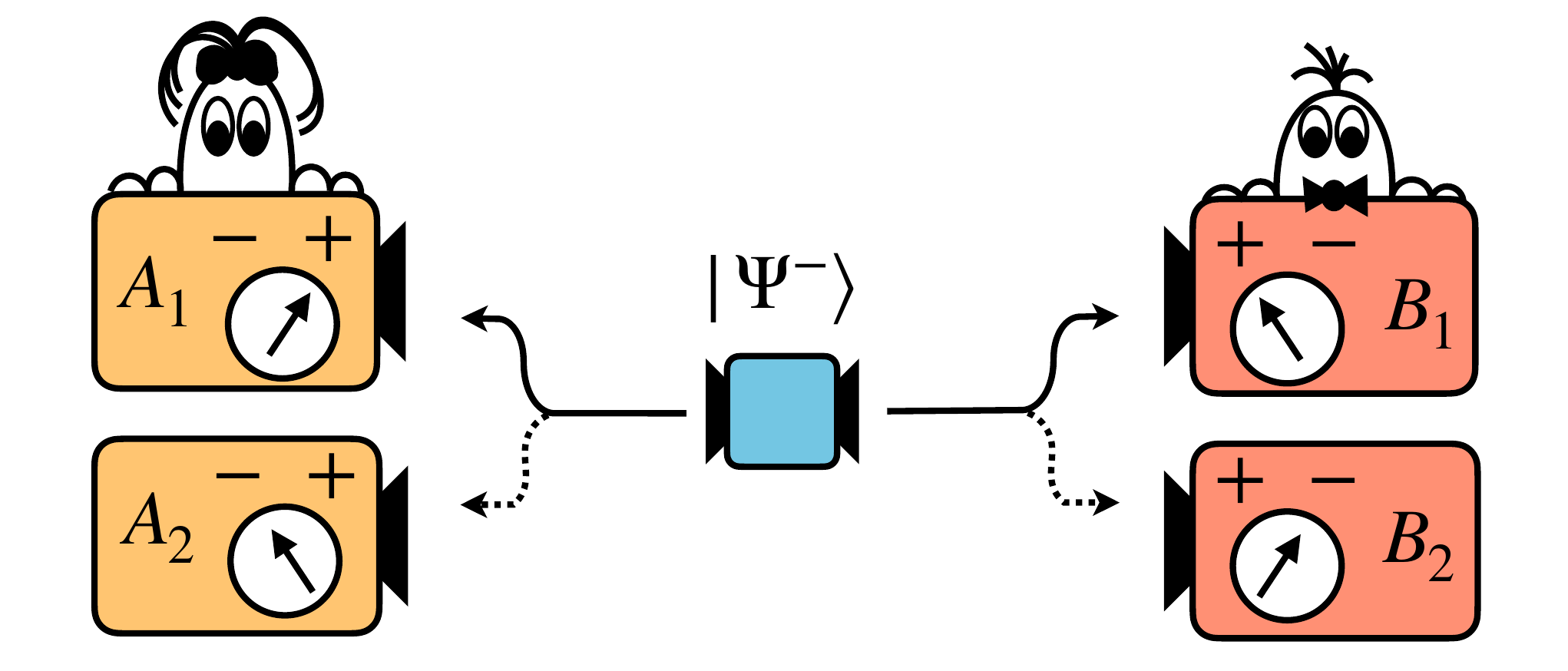}
\caption{Schematic picture of the Bell scenario. A source distributes two 
particles to two parties, named Alice and Bob. They perform measurements on
it, and the questions arises, whether the observed probability distribution
can be described by a local hidden variable model or not.}
\label{fig:bell-scenario}
\end{figure}

	To start, we discuss the relation between joint measurability and Bell nonlocality \cite{Bell:1964PHY, brunner14}. In this scenario, one considers two parties, Alice and Bob, and each 
of them performs some measurements $\{A_{a|x}\}$ and $\{B_{b|y}\}$, respectively
(see also Fig.~\ref{fig:bell-scenario}). Then, the question arises whether 
or not the observed probabilities $p(a,b|x,y)$ of the results $a,b$ for
the given inputs $x,y$ can be explained by a local hidden variable (LHV) 
model. This means that they can be written as
\begin{equation}
p(a,b|x,y) = \int \!\! d\lambda p(\lambda) \chi^A(a|x,\lambda) \chi^B(b|y,\lambda), 
\label{eq-lhv-model}
\end{equation}
where $\lambda$ is the hidden variable occurring with probability $p(\lambda)$
and $\chi^A(a|x,\lambda)$ and  $\chi^B(b|y,\lambda)$ are the response functions
of Alice and Bob, respectively. Note that here no reference to quantum mechanics
is made and no knowledge about the measurements on each side is assumed.

To start, let us explain why jointly measurable observables on 
Alice's side can never lead to Bell nonlocality \cite{fine82, wolf09}. For 
simplicity, we consider two measurements for Alice, $x=1,2$ with 
outcomes $a_1, a_2$, and analoguously for Bob. Since Alice's 
measurements are jointly measurable, Alice may perform the parent 
POVM  and directly obtain the probability distribution $p(a_1, a_2).$ If Bob 
performs one measurement $B_y$ simultaneously, they will observe the 
probability distribution $p(a_1, a_2, b_y| B_y)$. This has to obey
$p(a_1, a_2) = \sum_{b_y} p(a_1, a_2, b_y| B_y)$ independently of
$y$, since the correlations obey the non-signalling condition, i.e., 
Bob cannot send information to Alice by choosing his measurement.
Then, however, one may {\it define} a global probability distribution as
\begin{equation}
p(a_1,a_2, b_1, b_2) = \frac{p(a_1, a_2, b_1| B_1)p(a_1, a_2, b_2| B_2)}{p(a_1, a_2)}.
\end{equation}
Indeed, one can directly check that this obeys all the properties of
a probability distribution. The existence of such a distribution, however, 
implies already that the conditional distributions obey Eq.~(\ref{eq-lhv-model}), 
since the global distribution can always be expressed as a probabilistic mixture 
of deterministic assignments \cite{fine82}.

The question remains, whether any set of incompatible observables 
on Alice's side can lead to Bell nonlocality, if the underlying 
quantum state and Bob's measurements are properly chosen. For that, 
one needs to show that the correlations violate some suitable Bell
inequality, and connections between violations of Bell inequalities
and the degree of incompatibility have been observed quite early 
\cite{andersson2005,son2005}.

Interestingly, for the simplest scenario, where Alice and Bob have two
measurements each with two possible outcomes $\pm 1$, there is a direct 
connection between incompatibility and Bell nonlocality. In this case, 
the only relevant Bell inequality is the one by Clauser, Horne, Shimony
and Holt (CHSH) which reads \cite{Clauser:1969PRL, Clauser:1970PRL}
\begin{equation}
S = \mean{A_1 B_1} + \mean{A_1 B_2} +  \mean{A_2 B_1} - \mean{A_2 B_2} 
\leq 2.
\end{equation}
Here, $\mean{A_x B_y} = p(+,+|x,y) - p(+,-|x,y)- p(-,+|x,y) + p(-,-|x,y)$ denotes the expectation value of a correlation measurement.

The connection established by \citet{wolf09} uses the SDP formulation of
joint measurability. As shown in Section \ref{sec-SDP}, for two measurements 
$\{A_{a|x}\}$ with $x= 1,2$ the existence of a parent POVM $\{M_{\pm\pm}\}$ 
with four outcomes can be rephrased as a search for an effect $M_{++}$ obeying 
$
A_{+|1} + A_{+|2}- \openone \leq M_{++} \leq A_{+|x}
$
for $x=1,2$, see also Eq.~(\ref{eq-simple-sdp}). As mentioned, the search 
for such an effect $M_{++}$ can be formulated as a simple SDP, considered as 
a feasibility problem.

Given the SDP formulation in Eq.~(\ref{eq-simple-sdp}), one can consider the
dual SDP. As it turns out, this is directly linked to the CHSH inequality: The
additional variables of the dual problem can be viewed as a quantum state and
measurements on Bob's side and the CHSH inequality can be violated, if and only
if the SDP defined by Eq.~(\ref{eq-simple-sdp}) is unfeasible or, in other words, 
the measurements $\{A_{a|x}\}$ are incompatible.

For general scenarios the connection is, however, not so strict anymore. More
explicitly, \citet{bene18} presented a set of three measurements with 
two outcomes on a qubit, which are pairwise jointly measurable, but there 
is no common parent POVM for the entire set, so the triple is incompatible.
Then, it is shown for all quantum states and measurements on Bob's side, 
the resulting correlations are local in the sense of Eq.~(\ref{eq-lhv-model}).
This result holds for an arbitrary number of POVMs on Bob's side. Note that 
for the special case of dichotomic measurements on Bob's side, an analoguous
result was already shown by \citet{quintinox2016} and the case of an infinite 
number of measurements on Alice's side was considered by \citet{hirsch18}.

	\subsection{Quantum steering}
	\label{sec-steering}
	As we have mentioned in the above subsection, there are non-jointly measurable sets that can not break any Bell inequality for any quantum state. Here we review the results of \cite{quintino14,uola14,uola15,kiukas17} showing that for a slightly weaker form of correlations, the so-called quantum steering, incompatibility characterises exactly the sets of measurements that allow for the relevant non-local effect, which may be considered as a spooky action at a distance \cite{uola20review}.


The modern formulation of quantum steering is due to \citet{wiseman07}. In this formulation one party, say Alice, performs measurements $\{A_{a|x}\}$ in her local laboratory on a bipartite quantum state $\varrho_{AB}$. When asked to perform measurement $x$, she announces an output $a$ and the post-measurement state in Bob's laboratory is given by
\begin{align}\label{eq:stateassemblagedef}
\sigma_{a|x}=\text{tr}_A[(A_{a|x}\otimes\openone)\varrho_{AB}].
\end{align}
Assuming that the experiment is repeated many times and that Bob has access to a tomographically complete set of measurements, he can reconstruct these unnormalised states, also coined an assemblage $\{\sigma_{a|x}\}$. It is straight-forward to verify that the assemblage fulfils the condition of non-signalling, i.e., the operator $\sum_a\sigma_{a|x}:=\sigma_B=\text{tr}_A[\varrho_{AB}]$ is independent of the input $x$. 

The unsteerability of such an assemblage is associated to the existence of a local hidden state model. This is a model consisting of a local ensemble of states $\{p(\lambda)\varrho_\lambda^B\}$ on Bob's side, whose priors $p(\lambda)$ are updated upon learning the classical information $(a,x)$. In other words, an assemblage $\sigma_{a|x}$ is unsteerable if
\begin{align}\label{eq:LHSmodel}
\sigma_{a|x}=p(a|x)\sum_\lambda p(\lambda|a,x)\varrho_\lambda^B=\sum_\lambda p(\lambda)p(a|x,\lambda)\varrho_\lambda^B
\end{align}
and steerable otherwise. Here $p(a|x)=\text{tr}[\sigma_{a|x}]$ and the last equality follows from the fact that the hidden states are independent of the measurement choice, i.e. $p(x,\lambda)=p(x)p(\lambda)$, see \cite{uola20review} for a detailed interpretation of such models.

From Eq.~(\ref{eq:LHSmodel}) it is clear that separable states 
$\varrho_{AB}^\text{sep}=\sum_\lambda p(\lambda)\varrho_\lambda^A\otimes\varrho_\lambda^B$ can not lead to steerable assemblages. It is well-known, however, that entanglement is not sufficient for quantum steering \cite{wiseman07,quintino15}. For example, the Werner states \cite{werner89} and the isotropic states within a certain parameter regime provide examples of entangled states that have a local hidden state model for all measurements on Alice's side \cite{werner89,
barrett02, almeida07, wiseman07, nguyen20b, nguyenguehne2020}.

In spite of recent progress \cite{chau21}, the problem of deciding steerability of generic 
quantum states remains open. Still, a complete characterisation of the measurements that lead
to steering is known \cite{quintino14,uola14}. The first observation is that joint measurability 
of Alice's measurements leads to an unsteerable assemblage for any shared state. To see this, 
one can simply plug Eq.~(\ref{eq:jmdefinition}) into Eq.~(\ref{eq:stateassemblagedef}). Conversely, 
using the maximally entangled state $|\psi^+\rangle=\frac{1}{\sqrt{d}}\sum_{n=1}^d|n\rangle\otimes|n\rangle$ 
yields $\sigma_{a|x}=\frac{1}{d}A_{a|x}^T$, where $T$ denotes the transpose in the computational 
basis $\{|n\rangle\}$. Comparing Eq.~(\ref{eq:LHSmodel}) with Eq.~(\ref{eq:jmdefinition}) shows that 
a local hidden state model for $\{\sigma_{a|x}\}$ can be converted into a joint measurement 
of $\{A_{a|x}\}$ by denoting $G_\lambda=p(\lambda)d(\varrho_\lambda^B)^T$. We arrive at the 
following result:

\textit{Joint measurability of Alice's measurements leads to unsteerable assemblages 
for any shared quantum state. Conversely, for any set of incompatible measurements, 
there exists a shared state for which these measurements lead to steering.}

The above connection can be used to translate results on joint measurability to steering and vice versa \cite{chen16b,cavalcanti16,uola14}. For example, the incompatibility robustness of Alice's measurements is known to be lower bounded by so-called steering robustness of the corresponding assemblage \cite{chen16b,cavalcanti16}. Importantly, as steering verification does not assume Alice's measurements to be trusted, such lower bounds constitute semi-device independent bounds on measurement incompatibility, see also \cite{chen21} for further quantification techniques in the device independent setting. Moreover, \citet{uola14} showed that the characterisation of steerability of the isotropic state \cite{wiseman07} translates into a characterisation of the white noise robustness of all projective measurements in a given dimension.

There exist also two broader connections between the concepts of joint measurability and steering. First, the above result relies on the use of a pure maximally entangled state. To relax this, \citet{uola15} showed that the unsteerability of an assemblage $\{\sigma_{a|x}\}$ is equivalent to the joint measurability of the corresponding square-root or red``pretty good measurements". These are measurements that are known to give a good, but not always optimal, performance in discriminating the corresponding set of states \cite{Hausladen94} and they are known to relate to the information capacity of quantum measurements \cite{DallArno11,Holevo12}. They are given by $\sigma_B^{-1/2}\sigma_{a|x}\sigma_B^{-1/2}$, where $\sigma_B=\sum_a\sigma_{a|x}$ and a pseudo inverse is used when necessary. To see the more general connection between steering and measurement incompatibility, one can simply sandwich a local hidden state model in Eq.~(\ref{eq:LHSmodel}) with $\sigma_B^{-1/2}$, or sandwich a joint measurement in Eq.~(\ref{eq:jmdefinition}) with $\sigma_B^{1/2}$. We summarise this in the following.

\textit{A state assemblage $\{\sigma_{a|x}\}$ is steerable if and only if the corresponding 
pretty good measurements 
\begin{equation}
 B_{a|x}:=\sigma_B^{-1/2}\sigma_{a|x}\sigma_B^{-1/2}
\end{equation}
are incompatible.}

The above result gives a direct link between steering criteria and incompatibility conditions. 
For example, the incompatibility criteria of Section \ref{sec-MUR} were used to fully 
characterise the steerability of two-input two output qubit assemblages \cite{uola15}, cf. \cite{chenye17} for a detailed analysis, 
and it was shown that the incompatibility robustness of $\{B_{a|x}\}$ can be used to 
witness the entanglement dimensionality of the underlying bipartite state \cite{designolle21}. 
Another application of the connection was demonstrated in \cite{uola21}, where steerable states with a positive partial transpose \cite{moroder14} were used to construct incompatible qutrit measurements that are compatible in every qubit subspace. We note that similar results on incompatibility in subspaces were obtained by \citet{faedi21} 
using different techniques. Furthermore, the connection can be generalised for 
characterising so-called channel steering \cite{piani15a} via measurement 
incompatibility \cite{uola18}.

Going one step further, it was shown in \cite{kiukas17} that one can reformulate the steering problem of a state $\varrho_{AB}$ in the Choi picture. This gives a map between Alice's POVMs and the pretty good measurements on Bob's side. The channel associated to a shared state $\varrho_{AB}$ is given in the Heisenberg picture as
\begin{align}
\label{eq:ChoiChannel}
\Lambda_{\varrho_{AB}}^\dagger(A_{a|x})
=\sigma_B^{-1/2}\text{tr}_A[(A_{a|x}\otimes\openone)\varrho_{AB}]^T\sigma_B^{-1/2}=B_{a|x}^T,
\end{align}
where the transpose is in the eigenbasis of $\sigma_B$. We note that here a generalisation of the text-book Choi isomorphism is used, in which the fixed reduced state is $\sigma_B$ instead of the canonical maximally mixed state \cite{kiukas17}. It is straight-forward to see that the state $\varrho_{AB}$ is steerable if and only if the corresponding channel $\Lambda_{\varrho_{AB}}^\dagger$ does not break the incompatibility of some set of measurements.

The advantage of the channel approach is that it extends the connection between steering and non-joint measurability of Alice's measurements to the infinite-dimensional case and to POVMs with non-discrete outcome sets, and that it can unify seemingly different steering problems, such as the steerability of NOON-states subjected to photon loss and steerability in systems that have amplitude damping dynamics \cite{kiukas17}. Furthermore,  a similar map as in Eq.~(\ref{eq:ChoiChannel}) was used in the solution of the steering problem for two qubits \cite{chau21}.

	\subsection{Quantum contextuality}
	\label{sec-contextuality}
	Quantum contextuality refers to the fact that the predictions of quantum 
mechanics cannot be explained by hidden variable models which are 
noncontextual for compatible measurements \cite{Kochen:1967JMM}. In simple terms, measurements 
are compatible, if they can be measured simultaneously or in sequence 
without disturbance, and noncontextuality means that the model assigns
values to a measurement independent of the context, see also 
\cite{budroni21} for a recent review on the topic.

A remarkable fact is that contextuality can be proven regardless of the 
preparation of the quantum system. So, the connection between properties 
of quantum measurements and non-classical behaviour can be extremely strong 
in various scenarios. This is in stark contrast to distributed scenarios, 
where a properly chosen entangled state is required as a catalyst to 
harness the non-classical behaviour of the measurements. We review here 
the formal connection between the concept of joint measurability and 
contextuality for two different notions of contextuality.

\subsubsection{Kochen-Specker contextuality}

In the Kochen-Specker setup, the context is defined as a set of projective
measurements that can be performed simultaneously. One then asks whether 
a hidden variable model could explain the outcome statistics of all 
measurements, while assuming that the hidden variable assigns values 
independently of the context. This assumption leads to various contradictions
to quantum mechanics, e.g., in the so-called Peres-Mermin square \cite{Peres:1990PLA, 
Mermin:1990PRL}.

Kochen-Specker contextuality can be proven in a state-independent manner \cite{Cabello:2008PRL, 
yu12}. Hence, contextuality is a statement about measurements. The phenomenon of Kochen-Specker contextuality is based on the properties of the Hilbert space projections and it is indeed formulated for PVMs. On this level, the notions of joint measurability and non-disturbance reduce to commutativity. It is hence expected that non-commutativity is essential for the violations of the relevant classical models. However, as one needs also contexts for measurements, which requires compatibility, one needs to find the exact interplay between compatibility and incompatibility in order to reveal violations of Kochen-Specker non-contextuality. This structure was characterised by \citet{xu19} using graphs to represent the possible contextuality scenarios. In their graph representation, adjacent vertices represent compatible PVMs. The main result reads:

\textit{For a given graph, there exists a quantum realisation with PVMs producing contextuality if and only if the graph is not chordal.}

Here, chordality means that the graph does not contain induced cycles of size larger than three. Induced cycles are subgraphs with a set of vertices $S$ and edges $E$ such that the vertices $S$ are connected in a closed chain and, furthermore, every edge of the original graph that has both ends in $S$ is part of the subgraph. This implies especially that the simplest contextuality scenario requires four measurements.

\subsubsection{Spekkens contextuality}

The notion of operational non-contextuality asks whether one's measurement 
statistics can be reproduced by the means of an ontological model. Such a 
model assigns a distribution of ontological states $\lambda$ into each 
preparation procedure. This distribution is then classically post-processed. 
In short, for a preparation $P$ and a measurement $M$ with outcomes $\{a\}$, 
an ontological model reads
\begin{align}
\label{Eq:ontmodel}
p(a|P,M)=\sum_\lambda p(\lambda|P)p(a|M,\lambda),
\end{align}
where $p$ represents a probability distribution.

For quantum theory, where preparations are presented as density matrices 
and measurements as POVMs, such models can be constructed in many ways. 
For example, $(a)$ one can identify the space of ontological states as 
that of all mixed quantum states and define $p(\lambda|P)$ to be the point 
measure concentrated on $P$. Similarly, $(b)$ one can define a point 
measure on pure states and extend it to mixed states in a non-unique 
manner \cite{beltrametti95}. Also, $(c)$ one can identify quantum states 
with their eigendecompositions and choose $p(\lambda|P)$ to be eigenvalues 
and $\lambda$ to be the corresponding eigenprojectors.

In order to find contradictions with quantum theory, one needs to seek 
for meaningful restrictions of the ontological model \cite{spekkens05}. 
One 
possible restriction is to demand that operationally indistinguishable 
preparations are represented by the same distribution of ontological states. 
That is, if $P_1$ can not be distinguished from $P_2$, then 
$p(\lambda|P_1)=p(\lambda|P_2)$. This assumption is called preparation 
non-contextuality \cite{spekkens05}. An additional feature of these 
models is convex linearity, $p(\lambda|\mu P_1+(1-\mu)P_2) = \mu p(\lambda|P_1)+(1-\mu)p(\lambda|P_2)$ . 
Setting a similar restriction on indistinguishable measurements leads 
to the notion of measurement non-contextuality. It should be noted, however, 
that these assumptions are not obeyed by several known hidden variable models
\cite{Belinfante1973} and their physical relevance has been debated 
\cite{Ballentine:2014XXX}.

The above example models fit to these restrictions as follows. The 
models $(a)$ and $(c)$ are measurement non-contextual, but they break 
preparation non-contextuality in the sense that they are not convex 
linear. The model $(b)$ is measurement non-contextual, but breaks 
preparation non-contextuality in the sense that mixed states do not 
have a unique decomposition into pure states. Hence, measurement 
non-contextuality is not sufficient for contradicting quantum theory 
\cite{spekkens05}. Below we review the results of \cite{tavakoli19} 
showing that when all quantum state preparations are allowed, the notion 
of preparation non-contextuality together with its convex linearity 
feature are equivalent to joint measurability.

In the case of quantum theory, indistinguishable preparations are presented by the set of density matrices. Hence, the assumption of preparation non-contextuality restricts one to ontological models that depend only on the density matrix $\varrho$ and not the way $P$ it is prepared, i.e. $p(\lambda|P)=p(\lambda|\varrho)$. The assumption of convex linearity implies that for each $\lambda$ the map $p_\lambda(\varrho):=p(\lambda|\varrho)$ extends to a linear map from trace-class operators to complex numbers. As the dual of the trace-class is the set of bounded operators \cite{busch16}, one gets $p_\lambda(\varrho)=\text{tr}[\varrho G_\lambda]$ for all states $\varrho$ and for some positive operator $G_\lambda$. Noting that $\sum_\lambda p(\lambda|\varrho)=1$ for each state, it is clear that $\{G_\lambda\}$ forms a POVM. We summarise this in the following \cite{tavakoli19}.

\textit{Any jointly measurable set of POVMs leads to preparation non-contextual correlations for all input states. Conversely, the existence of a preparation non-contextual model for all quantum states implies joint measurability of the involved measurements.}

As a direct application, one sees that bounds on preparation contextuality translate to incompatibility criteria. For the explicit form of such witnesses, we refer the reader to \cite{tavakoli19}. 

We stress the fact that in the above result all quantum states are considered. In 
the experimental setting one does not have access to all possible states and, hence, 
the set of considered preparations is finite. In such scenario with a fixed set of 
states, there is no guarantee that an incompatible set of measurements would lead 
to preparation contextual correlations \cite{selby21}. It was further noted by \citet{selby21} 
that setting the additional restriction of measurement non-contextuality corresponds 
to a class of models that can be violated even by using compatible measurements.

	\subsection{Macrorealism}
	\label{sec-macrorealism}
	The notion of macrorealism challenges the classical intuition by asking the following question: Can one perform measurements in a way that does not disturb the subsequent evolution of the system? To formalise the concept, Leggett and Garg suggested to probe hidden variable theories that fulfil the assumptions of \textit{macroscopic realism} and \textit{non-invasive measurability}. In short, the first assumption amounts to the existence of a hidden variable $\lambda$ that carries the information about all measurements (despite them being performed or not) and the second assumption states that one can measure the system without disturbing the distribution of hidden variables. One should note that the second assumption is problematic, as it is not verifiable in the experimental setting. This is sometimes referred to as the \textit{clumsiness loophole} of macrorealism and there are various proposals for going around it \cite{leggett85,wilde11,knee12,Li12,george13,Emary13,budroni15,robens15,knee16,huffman17,emary17,Ku2020}. For this review, a central take on the problem is given by the measurement theoretical approach of \citet{uola19a}, which we explain in the following.

In a typical setting, one has $n$ time steps. On each step, one chooses either to perform or not to perform a measurement designated for that time step. Here, we concentrate exclusively on such scenarios. In this case, the assumptions of Leggett and Garg are equivalent to the fact that the resulting probability distributions are no-signalling in both directions \cite{clemente16}. In other words, all sequences of measuring and not measuring are compatible to one another under the act of marginalising. To give an example, consider the case $n=2$. Labelling the probability distributions as $p_i$ for a single measurement at time step $i=1,2$ and $p_{12}$ for the sequence, no-signalling in both directions requires
\begin{align}
\sum_a p_{12}(a,b)&=p_{2}(b),\\
\sum_b p_{12}(a,b)&=p_{1}(a)
\end{align}
for all outcomes $a$ and $b$.

Clearly, the second condition is satisfied by any physical distribution. However, requiring the first condition for all input states implies measurement compatibility in the form of non-disturbance. Importantly, the notion of non-disturbance involves the optimisation over all possible ways of performing the first measurement, i.e. optimisation over all instruments. This raises the following observation \cite{uola19a}, see also \cite{clemente15} where a connection between measurement compatibility and no-signalling conditions was discussed.

\textit{When all clumsiness caused by the lack of capabilities of the observer is removed, measurement incompatibility in the form of inherent measurement disturbance is the property that allows one to distinguish quantum theory from macrorealistic ones.}

For longer sequences of measurements, requiring the no-signalling constraints for all input states generates a more involved structure of non-disturbance relations. For example, in the case of three time steps, one requires non-disturbance in all pairs (in time order) and from the first measurement to the rest of the sequence. It is worth noting that one has to use the same instrument for a given time step in all conditions \cite{uola19a}. In other words, the first measurement has to be non-disturbing with respect to the second, the third and the non-disturbing sequence of the second and the third all with the same instrument. The structure arising from this generalised notion of non-disturbance is analysed in more detail in \cite{uola19a}, where also a resource theoretical take on the topic is discussed.

	\subsection{Prepare and measure scenarios}
	\label{sec-PM}
	In this Section we discuss the relevance of incompatible measurements in prepare-and-measure scenarios. More precisely, we will first review the advantage that incompatible measurements provide over compatible ones in state discrimination and state exclusion tasks. Then we will discuss the necessity of performing incompatible measurements in quantum random access codes, that is required to gain an advantage over classical random access codes. Lastly, we will review the role of incompatibility in distributed sampling scenarios.

\subsubsection{State discrimination and exclusion}
\label{sec:discr}

\begin{figure}[t]
\includegraphics[width=0.9\columnwidth]{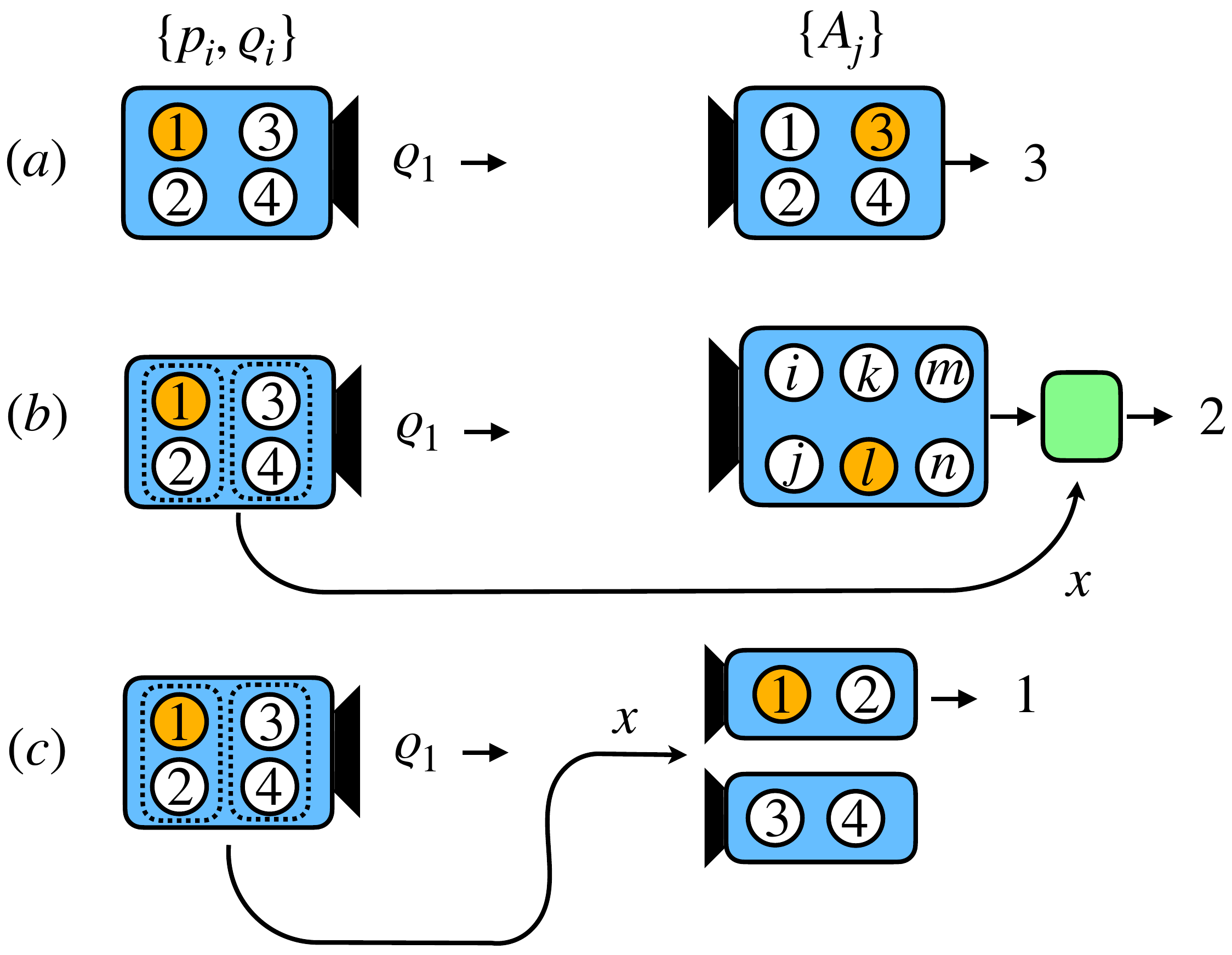}
\caption{Quantum state discrimination in different scenarios. 
(a)~In minimum-error state discrimination the task is to guess 
the correct input label of the state with high probability. 
(b)~In state discrimination with post-information the label $x$ 
is reviled after the measurement. Here, the additional information 
can only be used to post-process the classical measurement result. 
(c)~In state discrimination with prior-information the partition 
$x$ of the label is received before the state, and hence, 
a different measurement can be tailored to each subset $x$ of labels. Figure adapted from~\citet{Carmeli2018}. 
}
\label{fig:statediscrimination}
\end{figure}

A task that is important to quantum information theory, and in particular 
to quantum communication \cite{Helstrom1969, Holevobook} and quantum 
cryptography \cite{Bennett1984,Gisin2002Review} is \emph{minimum-error 
state discrimination} \cite{Helstrombook, Barnett09}, see Fig.~\ref{fig:statediscrimination}(a). There, one aims at 
correctly guessing the label of a state $\varrho_a$ that is randomly drawn 
from an ensemble $\mathcal{E}=\{p_a,\varrho_a\}_{a\in I}$, with known 
probability $p_a$. To be more precise, upon receiving a state, we perform 
a measurement of a POVM $\{A_a\}$, and guess the state to be $\varrho_a$, 
whenever we observe the outcome $a$. The  success in correctly guessing 
the label $a$ can be quantified by the \emph{probability of success} $
p_{\rm guess}(\mathcal{E},\{A_a\})=\sum_a p_a\tr[A_a\varrho_a]$. The maximum 
probability of success is simply obtained by maximizing over all measurements 
$p_{\rm guess}(\mathcal{E})=\max_{\{A_a\}}p_{\rm guess}(\mathcal{E},\{A_a\})$. 
We emphasize that this task is different from \emph{unambigous state discrimination}, 
where one is not allowed to make a wrong guess, but one is allowed to pass and not
give an answer at all \cite{Helstrombook, Barnett09}.

A similar but slightly different task is called \emph{minimum-error state-discrimination 
with post-measurement information} \cite{Ballester2008,Gopal2010}, see Fig.~\ref{fig:statediscrimination}(b). Suppose that the 
index set $I$ of the ensemble $\mathcal{E}$ is partitioned into non-empty disjoint 
sets $I_x$, such that $\bigcup_x I_x=I$ and that the label $x$ is revealed after 
the measurement of $\{A_a\}$ has been performed. This information 
cannot decrease the probability of guessing correctly the label $a$. However, the 
probability of success can increase if the label $x$ is revealed prior to the measurement, 
since one can tailor a separate measurement to each label $x$ individually, see Fig.~\ref{fig:statediscrimination}(c). Thus, 
one arrives at the conclusion that
$p_{\text {guess}}(\mathcal{E}) \leq p_{\text {guess}}^{\text {post}}(\mathcal{E}) \leq 
p_{\text {guess}}^{\text {prior}}(\mathcal{E})$. It was proven by \citet{Carmeli2018} 
that $p_{\text {guess}}(\mathcal{E}) = p_{\text {guess}}^{\text {post}}(\mathcal{E})
= p_{\text {guess}}^{\text {prior}}(\mathcal{E})$ if and only if the measurements that 
maximize the probability of success $p_{\text {guess}}(\mathcal{E})$ for each $x$ are 
jointly measurable. This also shows that joint measurability can be understood in terms of state discrimination games, namely, if the two scenarios in Fig.~\ref{fig:statediscrimination}(b) and (c) are indistinguishable, measurements are compatible.


The relation between incompatible measurements and state discrimination 
tasks can be made more precise by showing that whenever a set of measurements 
is incompatible, there exists an instance of a state discrimination task 
with prior information in which this set of measurements performs strictly 
better than any compatible one~\cite{skrzypczyk19,carmeli19a,uola19b,oszmaniec19,buscemi20}. 
More precisely, for any set of incompatible POVMs $\{A_{a|x}\}$, there exists 
a state discrimination task in which this set strictly outperforms any set 
of compatible measurements. The outperformance can be quantified by the 
incompatibility robustness $R_{\rm inc}(A_{a|x})$ and we have
\begin{equation}
\sup_{\mathcal{E}}\frac{p_{\text{succ}}(A_{a|x}, \mathcal{E})}{\max_{O_{a|x}\in JM} p_{\text{succ}}(O_{a|x},\mathcal{E})}=1+R_{\rm inc}(A_{a|x}).
\end{equation}
The state discrimination task can be derived from the optimal 
solution of the incompatibility robustness SDP in Eq.~\eqref{eq:JMRobustness}. The connection between state-discrimination games and the incompatibility robustness has been used to experimentally verify the incompatibility of two-qubit measurements in~\citet{Smirne2022}.
We note that a similar result was already known in the case of steering~\cite{piani15b},
Here, it was known that any steerable assemblage leads to a better performance in a suitably 
chosen \emph{sub-channel discrimination task} than any unsteerable assemblage. 

Comparing Eqs.~\eqref{eq:weight} and~\eqref{eq:robustness} one sees that the incompatibility
robustness and the incompatibility weight are very similar by their definition. Therefore, 
it is natural to ask whether the incompatibility weight has a similar interpretation in 
terms of state discrimination tasks. To that end one first needs to define \emph{minimum-error 
state exclusion tasks with prior-information}, also known as \emph{anti-distinguishability} 
\cite{Heinosaari2018}. Such tasks were first formalized by \citet{Bandyopadhyay2014} and in 
the context of the Pusey-Barrett-Rudolph argument against  a naive statistical interpretation 
of the wave function \cite{Pusey2012}.

The scenario in this task is similar as in minimum-error state discrimination, with the 
difference being that one aims to maximize the probability of guessing a state that was \emph{not} send, that is, one minimizes the probability $p_{\rm succ}(A_{a|x}, \mathcal{E})$ of guessing the state correctly.
Then one finds that \cite{uola19c} [see also \cite{Ducuara2020}]
\begin{equation}\label{eq:RatioMeasAs}
   \inf_{\mathcal{E}} \frac{p_{\rm succ}(A_{a|x},\mathcal{E})}{\min_{O_{a|x}\in \text{JM}}
   p_{\rm succ}(O_{a|x}, \mathcal{E})}= 1-W_{\rm inc}(A_{a|x}),
\end{equation}
where the optimization is performed over those sets $\{O_{a|x}\}$ for which the left-hand side is finite.

Finally, we note that the role of minimum-error state discrimination as a resource monotone and its connection to the robustness measure extends to the resource theory of single measurements (cf.~\citet{Skrzypczyk2019,Guff2021}).

\subsubsection{Quantum random access codes}


\begin{figure}[t]
\includegraphics[width=0.9\columnwidth]{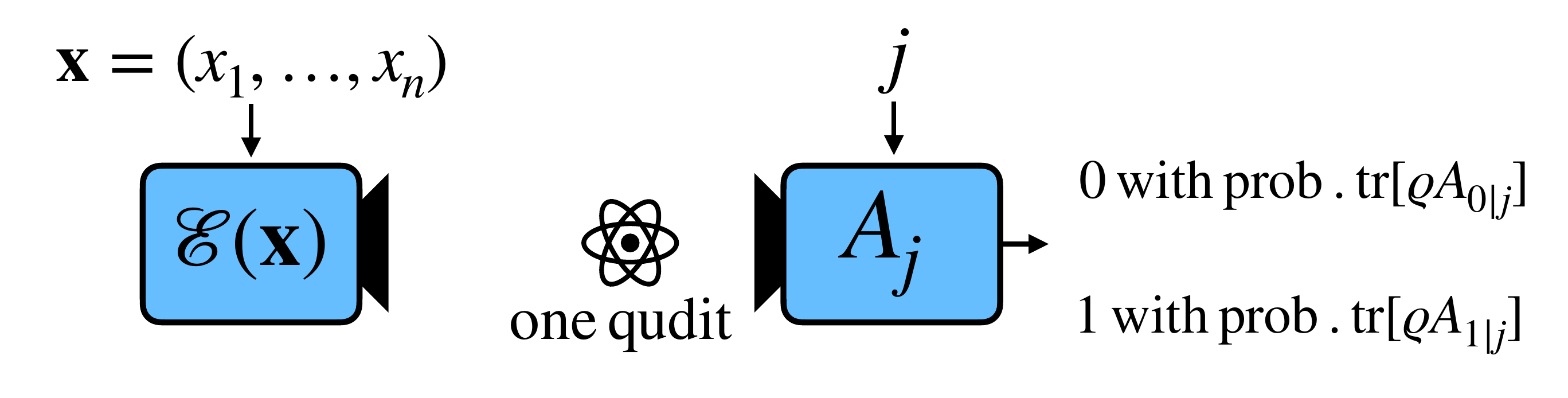}
\caption{Schematic view of quantum random access codes. A sender holds $n$ bits of information $\mathbf{x}$ and encodes it into a single qudit via the quantum channel $\mathcal{E}(\mathbf{x})$. The receiver wants to recover a random bit $x_j$ from $\mathbf{x}$ by performing a measurement $\{A_{x_j|j}\}$ on the qudit.}
\label{fig:QRAC}
\end{figure}


Random access codes (RACs) is an important class of classical communication 
tasks in which one party encodes a string of $n$ classical bits $\mathbf{x}=(x_1, . . . , x_n)$ 
into $m<n$ bits using some encoding strategy. Subsequently, the $m$ bits are communicated 
to a receiver. The task of the receiver is then to recover, with a high probability of 
success, a randomly chosen bit $x_j$ of the original string $\mathbf{x}$ using some 
decoding strategy. This is strongly related to the concept of information causality~\citet{Pawlowski2009}, which plays an important role in the foundations of quantum theory and in the problem of singling out quantum correlations from more general non-signalling correlations, cf.~\citet{Gallego2011}.

The idea of sending quantum states instead of classical information goes back 
to the work of~\citet{Wiesner1983}, where it was discussed under the name of 
\emph{conjugate coding}, and was later rediscovered by~\citet{Ambainis2002} 
in the field of quantum finite automata. In quantum random access codes 
(QRACs) the sender encodes the string of $n$ classical bits into a single 
$d-$level system using a CPTP map $\mathcal{E}(\mathbf{x})$, which is then called an $(n,d)$ QRAC. The 
decoding is done by performing a measurement $\{A_{x_j|j}\}$ depending 
on which bit $x_j$ was chosen to be recovered, see also Fig.~\ref{fig:QRAC}. 
The decoding was successful if the outcome is equal to the value of $x_j$. 
The average success probability is then given by
\begin{equation}
P_{\rm qrac}(A_1,\dots,A_n)=\frac{1}{nd^n}\sum_{\mathbf{x}} \tr[\mathcal{E}(\mathbf{x})(A_{x_1|1} + \dots + A_{x_n|n})],
\end{equation}
where $\mathcal{E}(\mathbf{x})$ is the encoding map and $A_{x_j|j}$ are the 
measurement effects. When optimized over states, the optimal average success 
probability is given by 
$\bar{P}_{\rm qrac}(A_1,\dots,A_n)=\frac{1}{nd^n}\sum_{\mathbf{x}} \Vert(A_{x_1|1} + \dots + A_{x_n|n})\Vert_\infty$, where $\Vert\cdot\Vert_\infty$ denotes the operator norm, and the measurements are useful if the average success probability exceeds 
the classical bound, i.e., if $\bar{P}_{\rm qrac}(A_1,\dots,A_n)>P_{\rm rac}^{n,d}$.

It was shown by \citet{carmeli19b} that only when incompatible measurements are used in the decoding step a QRAC performs better than its classical counterpart~(see also~\citet{FrenkelWeiner2015}). For $n=2$ the following results are known \cite{carmeli19b}:

\noindent
{\it
(1) For any compatible pair of $d$-outcome measurements $A_1$ and $A_2$ it holds that $\bar{P}_{\rm qrac}(A_1,A_2)\leq P_{\rm rac}^{2,d}$. The upper bound is tight.
}

This raises the question if \emph{all} incompatible measurements provide an advantage in QRACs. While this turns out not to be true in general, it is, however, true in the following two instances.

\noindent
{\it
(2) Let $A_1$ and $A_2$ be two sharp $d$-outcome measurements. Then, $\bar{P}_{\rm qrac}(A_1,A_2)\geq P_{\rm rac}^{2,d}$, with equality attained if and only if $A_1$ and $A_2$ are compatible.
}

\noindent
{\it
(3) Two unbiased qubit measurements $A_1$ and $A_2$ are incompatible if and only if they are useful for $(2,2)$-QRAC.
}

In the latter case, the result follows directly from the fact that the average success probability is a function of the Busch criterion in Eq.~(\ref{eq-jm-busch}). Furthermore, it was shown that there exist pairs of biased qubit observables that are incompatible, but nevertheless have $\bar{P}_{\rm qrac}(A_1,A_2)< P_{\rm rac}^{2,2}$, and thus do not provide an advantage over classical RACs.

In \citet{Hammad2020} quantum RACs have been demonstrated experimentally to quantify the \emph{degree of incompatibility} (see \citet{BLW2014}).  Another experimental implementation was reported in \citet{Foletto2020}.

\subsubsection{Distributed sampling}

Going beyond the state discrimination scenario from the previous section, other scenarios have been identified where incompatible measurements play an important role. \emph{Distributed sampling} refers to the task of Alice and Bob being able to sample from the probability distributions ${\Tr(\varrho_x B_{b|y})}_{b,x,y}$, where $\varrho_x$ is a quantum input of Alice and $y$ is a classical input of Bob~\cite{guerini19}. In case they share a perfect quantum communication channel Alice could send her input state to Bob, who can then perfectly sample from the desired probability distribution. When Alice's communication to Bob is restricted to classical information the most general strategy is that Alice performs a measurement $\{A_a\}$ on her input state $\varrho_x$ and sends her result to Bob. Bob then outputs a classical variable $b$ according to some response function $f(b|y,a)$. More precisely, the distributions that Alice and Bob can sample from are of the form $P(b|\varrho_x,y)= \sum \tr[\varrho_x A_a]f(b|y,a)$. One of the results of \citet{guerini19} is that a set of measurements is compatible if and only if the behaviour $\{\tr[\varrho_x B_{b|y}]\}$ admits a distributed sampling realization. Moreover, such a sampling task can be used to obtain lower bounds to the incompatibility robustness, and thus, quantify the degree of incompatibility of the implemented measurements.

\section{Further topics and applications}

In this Section we describe various topics related to quantum incompatibility. We start
with the potential resource theory of incompatibility. Then, we review various other
notions of ``incompatibility'', e.g., the incompatibility of channels, or other notions
for measurements, such as complementarity or coexistence. Finally, we shortly comment on
the problem of joint measurability for the infinite dimensional case, which is important
for understanding the incompatibility of position and momentum.

\label{sec-applications}

	\subsection{Resource theory of incompatibility}
	\label{sec-resource}
	Resource theories formalize the idea that certain operations or preparations 
require less resources than others. For instance, the preparation of separable states does not 
require any nonlocal operations, as local operations and classical communication (LOCC) are 
sufficient for their preparation. This is in stark contrast to entangled states, which require 
global operations for their preparation~\cite{horodecki09,guehne09}. Parts of entanglement theory 
may be seen as an example of a resource theory, in which separable states are deemed free, whereas 
entangled states are resourceful. A resource theory would then ask for sets of physically 
motivated monotones that decide if a transformation between two resourceful states is 
possible or not, e.g., the transformation of resources by free operations such as LOCC, or 
the distillation of highly resourceful states. The resourceful states can be used to accomplish 
some task, for instance, they can be used for teleportation or quantum key distribution.

In the case of measurement incompatibility, the compatible sets of POVMs are deemed 
resourceless and the resourceful measurements are the incompatible sets. To define 
meaningful resource monotones one first needs to establish a notion of free operations. 
\citet{Heinosaari2015b} considered as free operations pre-processing by quantum channels, 
i.e., every incompatibility monotone $I$ needs to satisfy $I[\Lambda(A_1),\Lambda(A_2)]\leq I[A_1,A_2]$, 
for all unital CPTP maps $\Lambda$. It was shown that in such a scenario a scaled 
violation of the CHSH inequality~\cite{wolf09} and the noise robustness are such monotones.  
In~\citet{guerini17,skrzypczyk19} classical post-processing was considered as a free operations. In that case, 
the relevant monotones need to be non-increasing under classical post-processing. It was 
shown by~\citet{skrzypczyk19} that the incompatibility robustness 
fulfils this property. Moreover, it was shown that state discrimination games with 
post-measurement information, forms a complete set of operationally meaningful monotones in the 
sense that a set of measurements $\{A_{a|x}\}$ can be transformed into a set 
$\{\tilde{A}_{a|x}\}$ by classical post-processing, if and only if 
$P_{\rm guess}(A_{a|x},\mathcal{E})\geq P_{\rm guess}(\tilde{A}_{a|x},\mathcal{E})$ 
for all state discrimination games $\mathcal{E}$.

From a physical point of view one is not necessarily constrained to choose 
between either pre- or post-processing. \citet{pusey15} considered CNDO operations as free operations and it was shown that for two binary projective measurements $\{B_{a|x}\}$ and $\{\tilde{B}_{a|x}\}$, the first one can be converted to the second one by CNDO if and only if they are unitary equivalent.
In~\citet{buscemi20}, both quantum 
pre-processing and conditional classical post-processing were considered. 
It was shown that $\{A_{a|x}\}$ can be transformed to $\{\tilde{A}_{a|x}\}$ via these operations
if and only if $\{A_{a|x}\}$ performs at least as good as $\{\tilde{A}_{a|x}\}$ in all discrimination games 
with post-information.

Finally,~\citet{Styliaris2019} studied the resource theory of measurement incompatibility relative to a basis. Here, probabilities arise from measurements on states which are diagonal in a fixed basis and one asks if the resulting probability distributions can be converted by classical post-processing. A connection between resource monotones and multivariate majorization conditions is shown.

	\subsection{Channel incompatibility}
	\label{sec-chaninc}
	Incompatibility can also be formulated for other quantum objects than measurements, especially 
for channels, i.e.\ completely positive trace-preserving maps. In short, we say that a set of 
quantum measurement devices (i.e.,\ a set of POVMs and a set of channels) is compatible if there 
is a single joint measurement process (represented by an instrument) that simultaneously realizes 
all the POVMs as well as all the channels in the set. These definitions follow the model set up in \citet{heinosaari17}; note that there the compatibility of a mixed set of POVMs and channels is seen as the compatibility of channels where the POVMs are replaced with appropriate measure-and-prepare channels. This generalizes the definition of joint measurability as we shall see. In the following, we will first give a precise definition of channel incompatibility, and then discuss two applications, the quantum marginal problem and information-disturbance trade-off.
	
\subsubsection{Formal definition of channel incompatibility}
\label{sec-chaninc-1}
To formalize the above idea, consider a set $\{A_{a|x}\}$ of POVMs on a system $A$ and a set of channels $\Lambda_y$ with the shared input system $A$ and possibly varying output systems $B_y$. Besides this, there are no further constraints on the relation between the set of POVMs and the set of channels. 

Imagine an instrument $\mathcal{I}=\{\mathcal{I}_{\vec{a}}\}$ whose input system is $A$ and output system is the composition of the systems $B_y$. Here, $\vec{a}$ denotes the vector $(a_x)_{x}$ that can encode 
all the outcomes of the POVMs $\{A_{a|x}\}$. Furthermore, we can denote by $E_{a|x}$ the set of all such vectors where the $x$th component is fixed to be $a$. Then, this instrument can be related to the set of POVMs and the set of channels from above in the following manner.

First, it may reproduce the POVMs, if ${\rm tr}[\varrho A_{a|x}]=\sum_{\vec{a}\in E_{a|x}}{\rm tr}[\mathcal{I}_{\vec{a}}(\varrho)]$ holds for all input states $\varrho$, all indices $x$ and all values $a$ of the $x$th POVM. Second, it may reproduce the channels in the sense that $\Lambda_y(\varrho)=\sum_{\vec{a}}{\rm tr}_{B_y^c}[\mathcal{I}_{\vec{a}}(\varrho)]$ for all  input states $\varrho$ and output labels $y$. Here, $B_y^c$ is the composition of all systems $B_{y'}$ where $y' \neq y.$ If such an instrument $\mathcal{I}$ exists, then we can say that the POVMs $\{A_{a|x}\}$ and the channels $\Lambda_y$ are contained in a single measurement setting. One can then also say that the POVMs and channels are {\it compatible}, otherwise they are {\it incompatible}. Similar definitions were introduced by \citet{HeMiRe2014}. 

One can immediately see that if the set of the output labels $y$ is empty (i.e., no 
channels are considered), then this definition corresponds to joint measurability of 
$\{A_{a|x}\}$ with the joint POVM $\{M_{\vec{a}}\}$ defined through 
${\rm tr}[\varrho M_{\vec{a}}]={\rm tr}[\mathcal{I}_{\vec{a}}(\varrho)]$ for all 
$\varrho$. On the other hand, if no POVMs are considered, i.e.,\ the set of labels 
$x$ is empty, the compatibility of channels $\Lambda_y$ is equivalent to the existence 
of a broadcasting channel (or joint channel) $\Gamma$ with the input system $A$ and 
whose output system is the composition of the systems $B_y$ such that 
$\Lambda_y(\varrho)={\rm tr}_{B_y^c}[\Gamma(\varrho)]$ for all $\varrho$ and $y$. 
This definition coincides then with those made in \cite{heinosaari17,haapasalo19b}. 
This notion of channel compatibility can also be generalized to channels that may 
share the input system only partly and also have overlapping output systems 
\cite{HsiLoAc2021}.

\subsubsection{Quantum marginal problem}

Marginal problems are compatibility problems of quantum states: Starting from given states in different subsystems one has to determine whether there is a global state from which all the subsystem states can be obtained as reduced states. This problem is also known as the $N$-representability problem \cite{Coleman63, ruskai1969n} and it remains a major problem in quantum chemistry \cite{national1995mathematical}. 

In mathematical terms, one has a collection $\mathcal{A}:=\{A_j\}_{j\in J}$ of quantum systems, where some subsets $X_i\subseteq\mathcal{A}$ with $i\in I$ are considered as subsystems. For each $i\in I$, there is a state $\varrho_i$ given on the system $X_i$. The marginal problem associated with this setting asks whether there is a global state $\varrho$ of the entire system $\mathcal{A}$ such that $\varrho_i$ is the reduced state of $\varrho$ for each $i\in I$, i.e.,\ $\varrho_i=\mathrm{tr}_{X_i^c}[\varrho]$. In principle, this problem can be formulated as an SDP, but often one has additional constraints, e.g., the 
global state $\varrho = \ketbra{\psi}{\psi}$ is required to be pure or bosonic or fermionic symmetries must be respected. For this case, systematic approaches using algebraic geometry \cite{klyachko2004quantum, klyachko2006quantum}, generalized Pauli constraints \cite{Castillo_et_al_2021},
or hierarchies of SDPs \cite{YuSiWyNgGu2021} have been developed, nevertheless, the problem remains hard. 

The central result of \citet{haapasalo19} tells that marginal problems and compatibility 
questions can be identified with each other through the generalized channel-state dualism 
defined by the fixed margin $\varrho_A$; see Subsection \ref{sec-steering} and Eq.~\eqref{eq:ChoiChannel} 
for the exact form of the map. See also Proposition 12 of \cite{Plavala2017} for the case of the canonical Choi-map, i.e.,\ the one where $\varrho_A$ in the maximally mixed state. In words, one can formulate: 

{\it A collection of $A \to B_i$ channels $\Lambda_i$ is compatible if and only if the 
marginal problem involving the corresponding Choi states has a solution.}

More explicitly, a tuple $\vec{\Lambda}=(\Lambda_i)_{i=1}^n$ of channels is compatible if and only if, for a full-rank state $\varrho_A$ which can be freely chosen, the marginal problem involving the Choi states $S_{\varrho_A}(\Lambda_i)$ of the channels has a solution. On the other hand, the marginal problem involving a given tuple $\vec{\varrho}=(\varrho_i)_{i=1}^n$ of states on systems $A\& B_i$ with the fixed 
$A$-margins $\tr_B[\varrho_i]=(\varrho_i)_A=\varrho_A$ has a solution if and only if the channels $\Lambda_i$ from $A$ to $B_i$ such that $\varrho_i=S_{\varrho_A}(\Lambda_i)$ are compatible. Technically, for this direction, we need $\varrho_A$ to be invertible but, as is pointed out in \cite{haapasalo19}, we are free to suitably restrict system $A$ to make $\varrho_A$ invertible simultaneously not effectively altering the original marginal problem, so this is not a real restriction. See also \cite{GiPlaSi2021} for details on similar results.

The above result enables the translation of results between the fields of compatibility and marginal problems. This was demonstrated in \cite{haapasalo19}, where entropic conditions for the solvability of the marginal problem \cite{carlen2013extension} were translated to necessary conditions for the compatibility of channels. Moreover, known conditions for the compatibility of channels \cite{haapasalo19b} were used to characterise the solvability of marginal problems involving Bell-diagonal states. Also solvability conditions for problems involving higher-dimensional qudit states with depolarizing noise were obtained, and the quantitative perspective of the connection was discussed.

Since measurements can be seen as quantum-to-classical channels, joint measurability questions can be recast as quantum marginal problems, too. Indeed, measurements $\{A_{a|x}\}_a$ can be identified with a measure-and-prepare channel $\Lambda_x$, $\Lambda_x(\varrho)=\sum_a{\rm tr}[\varrho A_{a|x}]\,\ketbra{a}{a}$, where the output system is a register with the orthonormal basis $\{\ket{a}\}_a$. It is 
easily seen that such measurements are jointly measurable if and only if the corresponding channels $\Lambda_x$ are compatible \cite{heinosaari17}. Moreover, a quick calculation shows that, for a full-rank input state in its spectral decomposition $\varrho_A=\sum_m \lambda_m\ketbra{m}{m}$ one finds the Choi states
\begin{equation}
 S_{\varrho_A}(\Lambda_x)=\sum_a\varrho_A^{1/2}A_{a|x}^T\varrho_A^{1/2}\otimes\ketbra{a}{a}
\end{equation}
where the transpose is taken w.r.t. the eigenbasis $\{\ket{m}\}_m$ of $\varrho_A$. Thus joint measurability questions can be identified with marginal problems involving block-diagonal states.

\subsubsection{Information-disturbance trade-off relations}
\label{subsubsec:inf-dist}

Here we discuss how the incompatibility between a single measurement and a specific quantum channel leads to an information-disturbance trade-off relation. Namely, we review the connection between the information gain in a measurement procedure and the inherent disturbance it causes on the system \cite{heinosaari13}.

Let us study the POVM $\{A_a\}_a$. To describe all the measurement processes describing this POVM, i.e.,\ all the instruments $\{\mathcal{I}_a\}$ such that $\tr[{\mathcal{I}_a(\varrho)}]=\tr[{\varrho A_a}]$ for all input states $\varrho$, let us fix a minimal Naimark dilation $\Delta$ for this POVM, see also Section \ref{subsubsec:Naimark}. This consists of an isometry $J$ of the input system to a larger dilation system and a PVM $\{P_a\}_a$ on the larger system,  such that $A_a=J^\dagger P_aJ$. It can be shown that any instrument $\{\mathcal{I}_a\}$ measuring $\{A_a\}$ is of the form $\mathcal{I}_a(\varrho)=\Phi(J\varrho J^\dagger P_a)$ for some channel $\Phi$ from the dilation system to the physical post-measurement system,
where $\Phi$ has to obey the additional constraint $\Phi(J\varrho J^\dagger P_a)=\Phi(P_a J\varrho J^\dagger)$ for all input states $\varrho$ and outcomes $a$ \cite{haapasalo14}. Using the condition on the channel $\Phi$, it follows that one can freely replace $\Phi$ with $\Phi\circ\mathcal{L}_{\Delta}$ where $\mathcal{L}_{\Delta}$ is the L\"{u}ders channel $\mathcal{L}_{\Delta}(\sigma)=\sum_a P_a\sigma P_a$. Recalling the definition of compatibility of POVMs and channels in Subsection \ref{sec-chaninc-1} above, this means that any channel $\Lambda$ compatible with single measurement $\{A_a\}$ is of the form $\Lambda=\Phi\circ\Lambda_{\Delta}$ for some channel $\Phi$ from the dilation system to the intended post-measurement system where $\Lambda_{\Delta}$ is the `maximal' channel $\Lambda_{\Delta}(\varrho)=\mathcal{L}_{\Delta}(J\varrho J^\dagger)$. This fact should be compared with the characterization given in Subsection \ref{subsubsec:Naimark} for the POVMs compatible with a given POVM which is completely analogical. 

Thus, the set $\mathfrak{C}_A$ of channels compatible with a fixed POVM $A=\{A_a\}$ has a very simple structure: these channels are all obtained by concatenating any channels to the maximal channel $\Lambda_{\Delta}$ determined by any Naimark dilation $\Delta$ of $A$. Using that any dilation can be connected to a minimal one with an isometry (see, e.g.,\ the construction of Subsection \ref{sec:retrieving} for this well known fact), it easily follows that, for any other dilation $\Delta'$, $\Lambda_\Delta$ and $\Lambda_{\Delta'}$ are equivalent in the sense that they are obtained from each other by channel concatenation. Thus, we may forget about the specific dilation and write $\Lambda_{\Delta}=:\Lambda_A$.

Using this simple structure of channels compatible with a fixed POVM, \citet{heinosaari13} proved 
a qualitative noise-disturbance trade-off relation: the noisier a POVM $A=\{A_a\}$ is, the larger 
the set $\mathfrak{C}_A$ of channels compatible with $A$ is. Specifically, given two POVMs $A=\{A_a\}$ 
and $B=\{B_b\}$ on the same system, there is a post-processing  $p(b|a)$ such that $B_b=\sum_a p(b|a)A_a$, 
if and only if $\mathfrak{C}_A\subseteq\mathfrak{C}_B$. 

Due to the simple structure of $\mathfrak{C}_A$ and $\mathfrak{C}_B$, the latter condition is 
equivalent to the existence of a channel $\Phi$ such that $\Lambda_A=\Phi\circ\Lambda_B$. Since 
a channel $\Lambda\in\mathfrak{C}_A$ describes the overall state transformation of the measurement 
of $A$, the larger the set $\mathfrak{C}_A$ (i.e.,\ the `higher' the maximal channel $\Lambda_A$) is, 
the less the measurements of $A$ can potentially disturb the system. 

Thus, we can interpret the above noise-disturbance relation in the following form: {\it the more 
informative (i.e.,\ less noisier) the measurement is, the more the measurement disturbs the 
system}. An extreme example is provided by the trivial POVMs where $A_a$ are all multiples of 
the identity operator whence $\mathfrak{C}_A$ is the set of all channels with the fixed input 
system, allowing complete non-disturbance. Naturally, these POVMs provide absolutely no 
information on the system being measured. Another example is the case of rank-1 POVMs 
where $A_a=\ketbra{\varphi_a}{\varphi_a}$. As these POVMs are maximal in the post-processing 
order, the sets $\mathfrak{C}_A$ are minimal. In fact, $\mathfrak{C}_A$ consists in 
this case of the measure-and-prepare channels of the form $\Lambda(\varrho)=\sum_a\tr[\varrho A_a]\,\sigma_a$ for some post-measurement states $\sigma_a$.

	\subsection{Further features of quantum measurements}
	\label{sec-further}
	Joint measurability is the main measurement theoretical notion 
in this review due to its various applications in quantum information 
science. Here, we discuss related concepts that have been used to grasp 
the counterintuive nature of quantum measurements.

\subsubsection{Simulability of measurements}\label{sec:sim}

There are various operationally motivated ways for relaxing the notion 
of joint measurability. One possible generalisation is to drop the 
assumption of having only one joint measurement. In other words, one can 
ask whether there is a simulation scheme that produces the 
statistics of $n$ measurements from $m<n$ POVMs. For instance, \citet{oszmaniec17} 
have defined the notion of measurement simulability as the existence of 
classical post-processings $p(a|x,y,\lambda)$ and pre-processings 
(or classical randomness) $p(y|x)$ such that
\begin{align}
A_{a|x}=\sum_{\lambda,y}p(y|x)p(a|x,y,\lambda)G_{\lambda|y},
\end{align}
where $x\in\{1,...,n\}$ and $y\in\{1,...,m\}$. Clearly, joint measurability 
corresponds to simulability with one POVM, i.e.\ $m=1$. This notion has also 
other interesting special cases such as measurements simulable with PVMs \cite{oszmaniec17}, 
measurements simulable with POVMs with a fixed number of outcomes \cite{kleinmann2016, guerini17,shi20}, 
and sets of measurements simulable with a given number of POVMs \cite{guerini17}. 

As examples, in \cite{oszmaniec17,hirsch17} projective simulability was used to improve the known 
noise thresholds for locality of two-qubit Werner state. Then, it was shown that 
any truly non-projective measurement, i.e. measurement not simulable with PVMs, 
provides an advantage in some minimum error state discrimination task over all 
projective simulable ones \cite{oszmaniec19,uola19b}. Finally, similar results 
for state discrimination with POVMs that can not be simulated with measurements 
having a fixed number of outcomes were reported in \cite{shi20}.

\subsubsection{Joint measurability on many copies}

Another extension of joint measurability may be introduced by using many 
copies of the given state \cite{carmeli16}. Of course, if one has two copies 
and two measurements, one can reproduce the statistics. In the simplest 
non-trivial scenario that deviates from joint measurability, one has a set of three POVMs $\{A_{a|x}\}$ and one is 
asked for a joint measurement on two copies of the original system together 
with post-processings for which
\begin{align}
    \text{tr}[A_{a|x}\varrho]=\sum_\lambda p(a|x,\lambda)\text{tr}[(\varrho\otimes\varrho)G_\lambda]
\end{align}
holds for all quantum states $\varrho$. As an example, it was shown by \citet{carmeli16} that three orthogonal noisy 
qubit measurements $A_{\pm|i}(\mu)=\frac{1}{2}(\openone\pm\mu\sigma_i)$ with $i=x,y,z$ have a 
joint measurement on two copies if and only if $0\leq\mu\leq\sqrt{3}/2$. 
In other words, there are triples with a two-copy joint measurement although each involved 
pair of measurements is incompatible. The structure of the incompatibility structures 
arising from compatibility on many copies were also analyzed in detail \cite{carmeli16}.

\subsubsection{Compatibility of coarse-grained measurements and coexistence}

Here we review special instances of incompatibility that raise from coarse-graining 
of measurements and discuss their relation to complementarity. Coarse-graining of 
a POVM corresponds intuitively to combining various POVM elements into a new one. 
More precisely, for a POVM $\{A_a\}$ we can consider disjoint subsets $E_i$ of the 
set of outcomes, and define the effects $A(E_i)=\sum_{a\in E_i}A_a$ and the coarse-grained 
POVM $\{A(E_i)\}_{i=1}^s$.  Especially, when $s=2$, the coarse-grained two-outcome POVM $\{A(E),\openone-A(E)\}$ is called 
a binarisation of $\{A_a\}$. 

As first notion of compatibility of coarse-grainings is that of \textit{coexistence}. 
This asks whether all yes-no questions, i.e. the set of \textit{all} possible 
binarisations $\{A_x(E_x),\openone-A_x(E_x)\}$ of given POVMs $\{A_{a|x}\}$, are jointly measurable with a {\it single} joint measurement. Here $E_x$ runs over all subsets of outcomes of the input $x$.
Note that for non-binary measurements, the set of all binarisations consists of more 
POVMs than the original set. 
This question can equivalently be reformulated as follows:
For any POVM $\{A_a\}$ one can define the range as the set of possible effects $A(E)$ in the notation
above.
Then, for a set of POVMs $\{A_{a|x}\}$ one can ask whether the union of their ranges is
contained in the range of a single POVM $\{G_\lambda\}$ \cite{Ludwigbook, busch16, Pekka03}; compare to \cite{haapasalo15}.  
Indeed, if
$
A_x(E_x)=\sum_{\lambda\in F}G_\lambda
$ 
then 
$
A_x(E_x)=\sum_\lambda p({\rm yes}|x,E_x,\lambda)G_\lambda
$ 
where $p({\rm yes}|x,E_x,\lambda)=1$ when $\lambda\in F$ and $0$ otherwise.
Note 
that joint measurability implies coexistence, but the inverse implication does 
not hold in general \cite{reeb13,pellonpaa14,uola21}. Naturally, similar notions 
can be defined also for more general coarse-grainings, but these have not been 
considered in the literature.

A closely related concept was proposed by \citet{saha20} as the notion of 
\textit{full complementarity}. This requires an incompatible pair $\{A_{a}\}$ 
and $\{B_{b}\}$ of POVMs to remain incompatible after arbitrary (non-trivial) 
coarse-grainings. The authors also defined the more general 
\textit{single-outcome 
complementarity} by demanding that for each pair $(a,b)$ the POVMs $\{A_{a},\openone-A_{a}\}$ 
and $\{B_{b},\openone-B_{b}\}$ are incompatible. Notably, here one 
is interested in the incompatibility of each coarse-grained pair 
separately. This is in contrast to coexistence. 

Note that if one would ask incompatibility of the binarizations for 
all single-outcome at once, i.e., with a single joint measurement, 
one would get a scenario that can be relevant for correlation experiments, in which one constructs 
multi-outcome measurements from two-outcome ones.
Albeit important, this last concept has not been analysed in the literature 
from measurement theoretical perspective.

\subsubsection{Complementarity}
\label{sec:complementarity}
Complementarity was defined to be equivalent to non-joint 
measurability in the works mentioned above \citet{saha20}, 
of which the notions of single-outcome complementarity and 
full complementarity are special cases. We note, however, 
that non-joint measurability is not the standard notion of 
complementarity, see e.g.~\cite{kiukas19}. Traditionally, 
a pair of POVMs $\{A_a\}$ and $\{B_b\}$ is called 
\textit{complementary} if all their measurements are mutually 
exclusive. Suppose that, for some outcomes $a$ and $b$, there 
is a positive operator $O$ such that 
\begin{equation}\label{comple}
\text{tr}[\varrho O]\le\text{tr}[\varrho A_a]
\quad\text{and}\quad 
\text{tr}[\varrho O]\le\text{tr}[\varrho B_b]
\end{equation}
for all states $\varrho$. If $O\ne 0$ then one gets nontrivial 
information on the measurement probabilities $\text{tr}[\varrho A_a]$ 
and $\text{tr}[\varrho B_b]$ in each state $\varrho$ for which the 
probability $\text{tr}[\varrho O]$ of the two-outcome measurement 
$\{O,\openone-O\}$ is not zero. Hence, the minimal requirement 
for complementarity is that Eq.~\eqref{comple} yields $O=0$ for 
all $a$ and $b$.

Clearly, complementary POVMs are incompatible since, for jointly 
measurable POVMs $\{A_a\}$ and $\{B_b\}$ with a joint POVM $\{M_{ab}\}$, 
Eq.~(\ref{comple}) holds for $O=M_{ab}$, which is non-zero for some $a$ and $b$.
Furthermore, the PVMs $A_a=|{\psi_a}\rangle\langle{\psi_a}|$ and 
$B_b=|{\varphi_b}\rangle\langle{\varphi_b}|$ related to mutually 
unbiased bases satisfy the minimal requirement. However, it is easy to see in the qubit case that after applying an arbitrarily small amount of white noise, the criterion is not fulfilled. Note that such measurements could still be complementary in the sense defined by 
\citet{saha20}.

One can pose stronger conditions on complementarity by writing the condition \eqref{comple} for general outcome sets $E$, $F$ with effects $A(E)=\sum_{a\in E}A_a$,  $B(F)=\sum_{b\in F}B_b$ instead of only singletons $\{a\}$, $\{b\}$. Of course, we must assume that the sets are such that $A(E)\ne\openone\ne B(F)$. In this way, one gets many different definitions of complementarity related to different choices of the sets in Eq.~\eqref{comple}. Especially, this works also for continuous POVMs in infinite dimensions. For example, the following pairs of POVMs related to the harmonic oscillator are complementary in the traditional sense: position-momentum, position-energy, momentum-energy, number-phase, and energy-time (which is essentially the number-phase pair) \cite{kiukas19}. \

\subsubsection{Retrieving measurements in the sequential scenario}
\label{sec:retrieving}

As we have discussed in the Section~\ref{subsec:commNDJM}, joint measurability 
and non-disturbance are not equivalent notions. However, jointly measurable 
pairs of POVMs do allow a sequential measurement scenario that recovers the 
data of both POVMs. Namely, instead of measuring a POVM $\{A_a\}$ (resp. $\{B_b\}$) 
on the first (resp. second) time step, one can measure $\{A_a\}$ on the first step 
and a retrieving measurement $\{\tilde B_b\}$ on the second step. In general, 
the retrieving measurement acts on a larger Hilbert space. It is also possible 
to remain in the original Hilbert space if one is allowed to choose the second measurement depending on the outcome of the first, i.e. one uses the output $a$ of the first measurement 
as an input for the second one.

For a detailed description, recall the notation from Section~\ref{subsubsec:Naimark} 
where a POVM $\{A_a\}$ was considered, with effects $A_a = \sum_{k=1}^{m_a} \ketbra{d_{ak}}{d_{ak}}.$
Instead of the additive Naimark extension in from Section~\ref{subsubsec:Naimark}
we consider now a product or `auxiliary' form Naimark dilation
\cite{preskill1998lecture} by choosing Hilbert spaces $\hil_n$ 
and $\hil_m$ with bases 
$\{e_a\}_{a=1}^n$ and $\{f_k\}_{k=1}^m$, $m\ge \max\{m_a\}$, taking
the tensor product  $\hil_n\otimes\hil_m$ (instead of $\hil_\oplus$), the projection
operators $P_a'=|e_a\rangle\langle e_a|\otimes\openone$, and the isometry
$J'=\sum_{a=1}^n\sum_{k=1}^{m_a}|e_{a}\otimes f_k\rangle\langle d_{ak}|$ \cite{Pello7}.
Note that by defining an isometry $W:=\sum_{a,k}|e_a\otimes f_k\rangle\langle e_{ak}|$ we get $WJ=J'$ and
$WP_a=P_a'W$ so, typically, the resulting Naimark extension $(\hil_n\otimes\hil_m,J',\{P_a'\})$ is 
not minimal (in which case, $J$ should be unitary and $nm=\sum_a m_a$, i.e.\ $m_a\equiv m$). However, for any joint measurement $\{M_{ab}\}$ of $\{A_a\}$ and 
$\{B_b\}$ there exists a (possibly nonunique) POVM $\{\tilde B_b\}$ of $\hil_m$
such that $M_{ab}={J'}^\dagger(|e_a\rangle\langle e_a|\otimes\tilde B_b)J'$ 
\cite{Pello7}.
Especially, we obtain the sequential measurement interpretation of the joint measurement.
We can write $\text{tr}[M_{ab}\varrho]=\text{tr}[\mathcal I_a(\varrho)\tilde B_b]$, where 
we have expressed an instrument $\{\mathcal I_a\}$ in the Stinespring form, $\mathcal I_a(\varrho):=\text{tr}_{\hil_n}[J'\varrho{J'}^\dagger(|e_a\rangle\langle e_a|\otimes\openone)]$. 

If we choose $m\ge d$ and define isometries $J_a:=\sum_{k=1}^d|e_a\otimes f_k\rangle\langle\varphi_{ak}|$ we obtain an alternative form
$\mathcal I_a(\varrho)=\Lambda_a(\sqrt{A_a}\varrho\sqrt{A_a})$ where 
$\Lambda_a(\varrho):=\text{tr}_{\hil_n}[J_a\varrho{J_a}^\dagger]$ is a quantum channel. From here one 
can find the retrieving measurements on the original system by setting 
$\tilde B_{b|a}:=\Lambda_a^\dagger(\tilde B_b)$. Clearly for each $a$ this forms a POVM. 
One further sees that the L\"uders instrument is the least disturbing in the sense that 
after it, the data of any POVM jointly measurable with $\{A_a\}$ can be retrieved. This fact has also been found in \cite{heinosaari15b} where the minimal disturbance property, also called as universality, is associated to the channel $\Lambda_A$ of Subsection \ref{subsubsec:inf-dist}.

\begin{center}
\begin{figure}
\includegraphics[width=0.8\columnwidth]{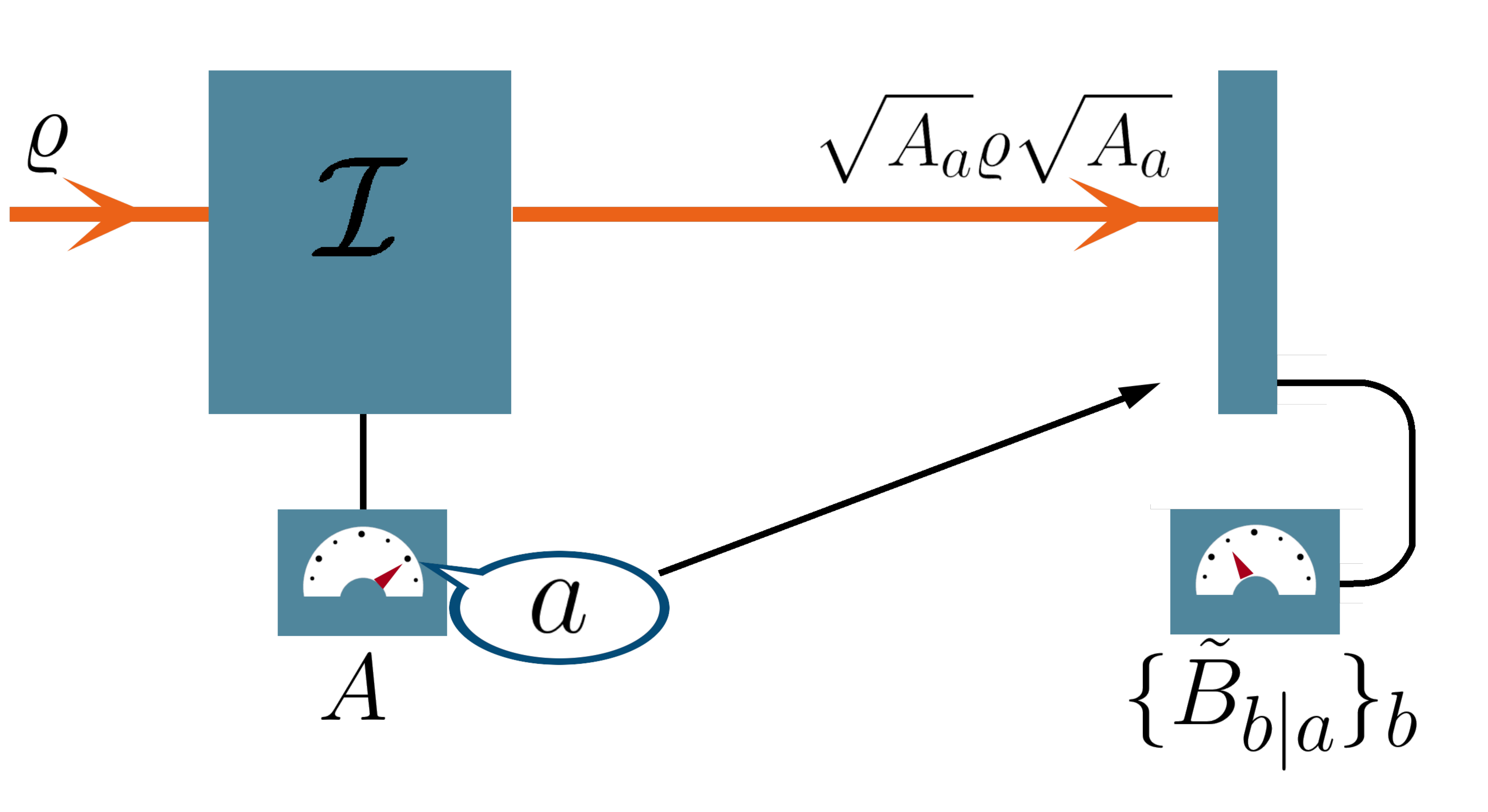}
\caption{\label{fig:sequential} Illustration for a joint measurement (a parent measurement) for the target POVMs $A$ and $B$: As long as $A$ and $B$ are jointly measurable, by first performing the L\"{u}ders measurement of $A$ and then, conditioned by the outcome $a$ obtained in the $A$-measurement, measuring a conditional POVM $\{\tilde{B}_{b|a}\}_b$ after the $A$-measurement, this sequential scheme realizes a joint measurement of $A$ and $B$.}
\end{figure}
\end{center}

\begin{center}
\begin{figure}
\includegraphics[width=0.8\columnwidth]{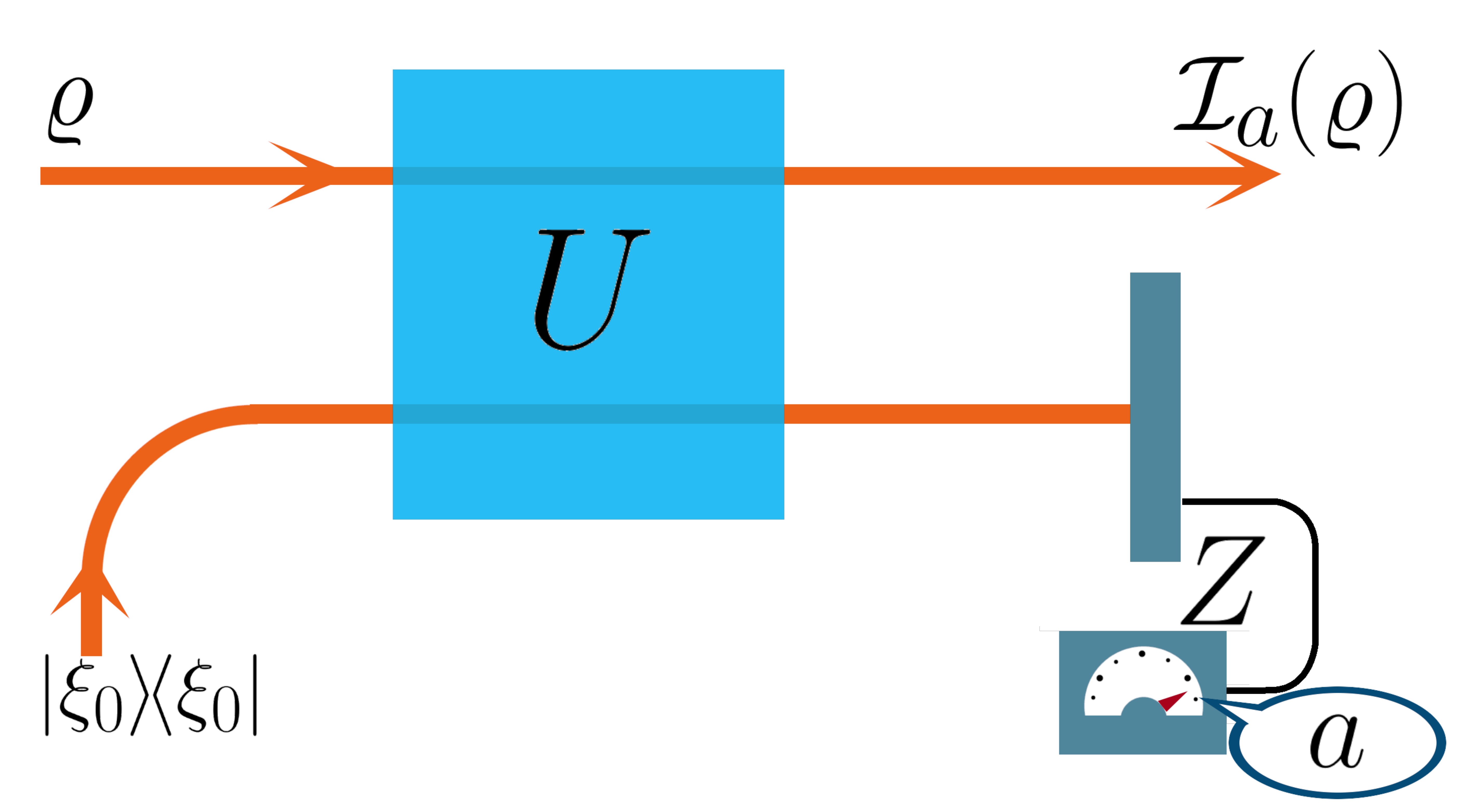}
\caption{\label{eq:measurementscheme} The `auxiliary' Naimark dilation has a particular form which enables the following physical interpretation of an instrument $\mathcal{I}$ measuring the target POVM $A$ \cite{ozawa}: The system state is coupled with a probe (the auxiliary system in the dilation) in initial state $|\xi_0\rangle\langle\xi_0|$. A unitary evolution mediated by the unitary $U$ then takes place followed by a sharp pointer measurement $Z$ of the probe after which the probe is detached from the system.}
\end{figure}
\end{center}

For a further physical interpretation, consider the case $m=d$. Then, we can identify the 
system's Hilbert space with $\hil_m$ and `extend' the isometry $J'$ to a unitary operator 
$U$ on $\hil_n\otimes\hil_m$ via $U(\ket{\xi_0}\otimes\ket{\psi}):=J'\ket{\psi}$ for 
$\ket{\psi}\in\hil_m$, where $\ket{\xi_0}\in\hil_n$ is some fixed ancilla's  unit vector 
(the so-called `ready' state). Now $\langle\hil_n,\{Z_a\}_a,\ket{\xi_0},U \rangle$ 
is called a measurement scheme (or measurement model) of $\{A_a\}$; especially $\{Z_a\}$, $Z_a=|e_a\rangle\langle e_a|$, is the pointer PVM, see e.g.\ \cite{busch96}.

It has a clear physical meaning: Before the measurement, the initial state of the 
compound system is $|\xi_0\rangle\langle\xi_0|\otimes\varrho$ since one assumes 
that the probe (ancilla) and system are dynamically and probabilistically independent 
of each other. Then the measurement coupling $U$ transforms the initial state into 
the final (entangled) state $S_\varrho:=U(|\xi_0\rangle\langle\xi_0|\otimes\varrho)U^\dagger=J'\varrho J'^\dagger$ 
which determines the subsystems' final states $\text{tr}_{\hil_m}[S_\varrho]$ and  
$\text{tr}_{\hil_n}[S_\varrho]$. 

The probability reproducibility condition $\text{tr}[\varrho A_a]=\text{tr}[S_\varrho(Z_a\otimes\openone)]=\langle e_a|\text{tr}_{\hil_m}[S_\varrho]|e_a\rangle$ guarantees 
that the measurement outcome probabilities are reproduced in the distribution of 
the pointer values in the final probe state. The state 
$S_\varrho^a:=\text{tr}[\varrho A_a]^{-1}(Z_a\otimes\openone)S_\varrho(Z_a\otimes\openone)$ can be interpreted as a conditional state 
under the condition $|e_a\rangle\langle e_a|\otimes\openone$ \cite{Cassi}. One obtains the subsystem 
states $\varrho_a:=\text{tr}_{\hil_n}[S_\varrho^a]$ and $|e_a\rangle\langle e_a|$, where 
the last one is the state of the probe after the interaction on condition 
$Z_a$. The sequential interpretation of the joint probability 
distribution,
\begin{equation}
\text{tr}[M_{ab}\varrho]=\text{tr}[S_\varrho(Z_a\otimes\tilde B_b)]=\text{tr}[\varrho A_a]\text{tr}[\varrho_a\tilde B_b],
\end{equation}
shows that the states $\varrho_a$ give the probabilities for any subsequent measurement on the system.
{In addition, any joint POVM can be expressed in the tensor product form.}

\subsection{Compatibility of continuous POVMs}

So far, we have explained the notions of incompatibility for the 
case of POVMs in finite-dimensional spaces with a finite set of 
outcomes. But, as we have mentioned already in the introduction, 
in the entire research program the case of position and momentum
observables played an outstanding role \cite{heisenberg1925, 
bornjordan1925}. So, we will finally explain some basic facts
about joint measureability for the case of `continuous' POVMs,
to demonstrate that most of the questions studied in this review 
are relevant also in the continuous case. However, as the continuous 
case is not our main focus, we only briefly outline how compatibility 
is defined in this setting and mention some generalizations of the 
results presented above.

Let us study a set  $\{A_x\}$  of POVMs labeled by $x$, where each POVM has a 
`continuous' value space $\Omega_x$. Here the allowed events or outcomes are 
measurable subsets of $\Omega_x$.
These POVMs operate in a possibly infinite-dimensional Hilbert space 
$\mathcal{H}$. We say that the set $\{A_x\}$ is jointly measurable 
if and only if there is a POVM $G$ 
and conditional probability measures $p(\cdot|x, \lambda)$ 
for which
\begin{equation}
A_x(E)=\int p(E|x, \lambda)\,dG(\lambda)
\end{equation}
for all measurable $E \subseteq \Omega_x$. Note that in Eq.~\eqref{eq:contPOVMdensity} 
below we present for any $A_x$ a density with respect to a probability measure,
giving a concrete way of evaluating the above operator integral. Joint measurability can be equivalently defined by requiring that there is a parent POVM $M$ from which $A_x$ can be obtained as margins just as in the discrete case.

Also in the continuous case, PVMs are jointly measurable only if they commute. From this we immediately see that the canonical position and momentum are not jointly measurable; see also our subsequent discussion on the quadrature observables. Position and momentum are also maximally incompatible in the sense that the addition of the maximum amount of trivial noise is required to make them compatible \cite{HeSchuToZi2014}. This quantification of incompatibility is similar to the incompatibility random robustness with the addition that the noise can consist of any POVMs whose effects are multiples of the identity operator.

In our previous discussions, e.g., in Sections \ref{subsubsec:Naimark} and \ref{sec:retrieving},
the Naimark extension played an important role, so let us explain this also for the continuous
case. First, a general POVM $A$ of the possibly infinite dimensional Hilbert space $\mathcal H$ 
with a basis $\{|n\rangle\}$, can be written as
\begin{align}
\label{eq:contPOVMdensity}
A(E)=\int_E \sum_{k=1}^{m_a}|d_{ak}\rangle\langle d_{ak}|d\mu(a)
\end{align}
where $E$ is a measurable subset of outcomes, $\mu$ is a probability (or positive) distribution 
on the outcome space, the integral runs over all $a \in E$, 
and $|d_{ak}\rangle$'s are generalized vectors \cite{Pello3}. 
By defining $\mathcal H$--valued wave functions $|\psi_n\rangle$ as maps $a\mapsto\ket{\psi_n(a)}=\sum_{k=1}^{m_a}\langle d_{ak}|n\rangle|k\rangle
$
and an isometry 
$
J=\sum_n |\psi_n\rangle\langle n|
$
one obtains a minimal Naimark dilation
\begin{align}
A(E) & =J^\dagger P(E) J = \sum_{m,n}\langle\psi_m|P(E)|\psi_n\rangle|m\rangle\langle n| \nonumber\\
&= \sum_{m,n}\int_E\langle\psi_m(a)|\psi_n(a)\rangle d\mu(a)|m\rangle\langle n|,
\end{align}
where $P$ is the `generalized position PVM' of the wave function 
space (a direct integral) defined via $(P(E)\psi)(a)=\psi(a)$ when $a\in E$ and 0 otherwise. 

If a POVM $B$ is jointly measurable with $A$, then $B(F)=J^\dagger \tilde B(F) J$, where
the POVM $\tilde B$ commutes with $P$ [i.e., it is of the form
$
(\tilde B(F)\psi)(a)=\tilde B_a(F)\psi(a)
$
where $\tilde B_a$ is a POVM acting in the $m_a$--dimensional 
subspace of $\mathcal H$ \cite{pellonpaa14}]. Especially, if each 
multiplicity is given by $m_a=1$, meaning that $A$ is of rank 1 \cite{Pello9}, 
then the operators $\tilde B_a(F)$ are conditional probabilities, say $p(F|a)$, 
and we have $B(F)=\int p(F|a)d A(a)$, showing that $B$ is a classical 
postprocessing or smearing of $A$. This is in line with the finite-dimensional 
result in Section~\ref{subsubsec:Naimark}. Indeed, in the discrete case $\mu$ is the counting measure so all integrals reduce to sums: $A(E)=\sum_{a\in E}A_a$ where $A_a=\sum_{k=1}^{m_a}|d_{ak}\rangle\langle d_{ak}|=\sum_{m,n}\langle\psi_m|P_a|\psi_n\rangle |m\rangle\langle n|= \sum_{m,n}\langle\psi_m(a)|\psi_n(a)\rangle |m\rangle\langle n|$ and $P_a=|e_a\rangle\langle e_a|\otimes\openone$, $\psi_n=\sum_a |e_{a}\otimes\psi_n(a)\rangle=\sum_{a,k} |e_{a}\rangle\otimes\langle d_{ak}|n\rangle |k\rangle$.
Now the wave function space is just a direct sum $\hil_\oplus=\bigoplus_a\hil_a$ where 
each $\hil_a$ is spanned by the vectors $|e_a\rangle\otimes|k\rangle$, $k=1,2,\ldots,m_a$. Furthermore,
operators $\tilde B(F)$ above are of the `diagonal block form', i.e., any $\tilde B_a(F)$ is an operator of $\hil_a$.

For example, in the case of a covariant phase POVM \cite{Holevobook,busch16}
\begin{equation}
\Phi(E)=\sum_{m,n=0}^\infty \langle\eta_m|\eta_n\rangle\int_E e^{i(m-n)\theta}\frac{d\theta}{2\pi}|m\rangle\langle n|,\; E\subseteq[0,2\pi),
\end{equation}
where $\ket{\eta_n}$'s are unit vectors which span a $d$--dimensional space, one sees that $m_\theta=d$, $\ket{\psi_n(\theta)}=\ket{\eta_n} e^{-in\theta}$, and 
any jointly measurable POVM of $\Phi$ can be written as 
\begin{equation}
B(F)=\sum_{m,n=0}^\infty\int_0^{2\pi}  \langle\eta_m|\tilde B_\theta(F)|\eta_n\rangle e^{i(m-n)\theta}\frac{d\theta}{2\pi}|m\rangle\langle n|.
\end{equation}
For instance, if $\tilde B_\theta=\tilde B$ does not depend on $\theta$ we get a 
smeared number observable
$
B(F)=\sum_{n=0}^\infty \langle\eta_n|\tilde B(F)|\eta_n\rangle|n\rangle\langle n|.
$
Or, if $\Phi$ is the (rank-1) canonical phase \cite{Pello2}, i.e.\ any 
$\ket{\eta_n}=|0\rangle$ and $d=1$, the above formula reduces to $B(F)=p(F)\openone$, with 
$p(F)=\langle 0|\tilde B(F)|0\rangle$, which is a trivial smearing of both the canonical 
phase and the sharp number. In conclusion, one cannot measure the canonical phase and 
the number together [actually they are complementary observables \cite{Pello10}],
but after suitable smearings they become compatible.

Similarly, the rotated quadratures $Q_\theta=Q\cos\theta+P\sin\theta$ are of rank 1, 
so that they are not jointly measurable (here $(Q\psi)(x)=x\psi(x)$ and 
$(P\psi)(x)=-i\hbar d\psi(x)/dx$ or, in Dirac's notation, $\bra{x} Q \ket{\psi} = x \braket{x}{\psi}$ and $\bra{x} P \ket{\psi} = -i \hbar \partial_x \braket{x}{\psi}$,
are the position and momentum operators). However, their smeared versions have joint measurements, see, e.g., \cite{busch16} 

Finally, many of the results and interpretations of joint measurability presented in 
this review generalize to the continuous setting. For example, the results of 
Section~\ref{sec:discr} on the advantage that incompatible measurements provide 
in state discrimination tasks can be generalised to the continuous variable setting 
as done by \citet{kuramochi20}, where the incompatibility robustness still quantifies 
the advantage. Here, it is worth noting that the discrete version of the result relies on the SDP formulation of joint measurability, but such formulation does not exist for the continuous case. This leads to the use of more involved techniques through limit procedures, see \citet{kuramochi20} for details. Moreover, the connection between steering and measurement incompatibility 
detailed in Subsection \ref{sec-steering} is generalized to continuous measurements and 
infinite-dimensional Hilbert spaces by \citet{kiukas17}. Furthermore, the $W$-measure of Section~\ref{sec:wmeasure} bears similarity to the continuous variable $s$-parametrised 
quasi-probability distributions \cite{cahill}. The connection between these distributions and joint measurability was studied in \cite{rahimi-keshari21,Pello8}. In \cite{Pello8} it was shown that between the Wigner and 
the $Q$-function, these distributions relate to operator-valued measures with possibly non-positive semi-definite 
elements, whose marginals are the smeared position and momentum measurements. In the special 
case of the $Q$-function, one gets a joint measurement for noisy position and 
momentum measurements. However, obviously not all of the results obtained for discrete POVMs can be extended to the continuous setting. For example, the fact that for discrete POVMs $A$ and $B$ we have that $A$ is a post-processing of $B$ if and only if any channel compatible with $B$ is also compatible with $A$ has not yet been established in the continuous case, although it makes operational sense as the core of the information-disturbance trade-off presented in Section \ref{subsubsec:inf-dist}.


\subsection{Glossary}
A glossary summarizing different types of measurements and their definitions can be found in Table~\ref{tab:glossary}.

\begin{table*}[t!]
\begin{tabularx}{\textwidth}{lX}
\hline
\textbf{Term} & \textbf{Definition} \\ \hline
\rowcolor[HTML]{EFEFEF} 
Compatible measurements & Can be simultaneously classically post-processed from a single measurement, see Eq.~\eqref{eq:jmdefinition}. \\
Incompatible measurements & Measurements that are not compatible. \\
\rowcolor[HTML]{EFEFEF} 
Joint measurement & The measurement from which compatible ones can be post-processed via Eq.~\eqref{eq:jmdefinition}. \\
Parent (or mother) POVM & A joint measurement which is of the marginal form in Eq.~\eqref{eq:MargJM}. \\
\rowcolor[HTML]{EFEFEF} 
Complementary measurements & 
Two POVMs are complementary if sufficiently many pairs of their effects are mutually exclusive. \\
Unbiased measurement & Produces a uniform probability distribution when measured on the maximally mixed state. \\
\rowcolor[HTML]{EFEFEF} 
Non-disturbing measurements & A POVM is said to be non-disturbing with respect to another POVM if there exists a sequential implementation in which neglecting the outcome of the first measurement does not affect the statistics of the second measurement. \\
Pretty good measurement & Preforms pretty good (but not optimal) in state discrimination tasks where states are roughly of equal probability and almost orthogonal. \\
\rowcolor[HTML]{EFEFEF} 
Retrieving measurement & A measurement that is used to implement compatible POVMs in a sequential order, see Sec.~\ref{sec:retrieving}. \\
Simulable measurements & Generalized notion of measurement compatibility, where statistics of measurements are simulated by fewer measurements, see Sec.~\ref{sec:sim}. \\
\rowcolor[HTML]{EFEFEF} 
Coarse-grained measurement & Any measurement that is obtained by binning the outcomes of a measurement. \\ \hline
\end{tabularx}
\caption{Glossary summarizing different types of measurements that where discussed in this article.}
\label{tab:glossary}
\end{table*}

\section{Conclusion}
\label{sec-conclusion}
The puzzling properties of quantum measurements have sparked intense
discussions among scientists since nearly hundred years. This led to 
many interesting research results, and it took some time until key 
concepts of quantum measurement theory, such as the notion of POVMs and
the notion of joint measurability have emerged and found widespread 
applications. In this review, we explained the incompatibility of 
measurements, highlighting the connections to information processing. 

One may hope that concepts like POVMs, instruments, and joint measurability 
will become standard knowledge on quantum mechanics in the physics community 
in the future. This is motivated by the fact that an operational view on quantum 
mechanics combined with elements of information theory is becoming more and more 
standard in physics. For instance, many new textbooks and university courses favour 
this viewpoint, showing that the way of teaching and understanding quantum mechanics 
is changing. In addition, also novel applications of joint measurability
may be found. 

There are several interesting open problems connected to the incompatibility 
and joint measurability of generalized measurements and the following list 
gives a small selection:
\begin{itemize}
\renewcommand\labelitemi{--}
\item It has been shown that not all incompatible measurements can lead 
to Bell nonlocality. So, which additional properties of measurements are 
required for this?

\item As we have seen, incompatibility can be quantified by different 
figures of merit. The question arises, which are the most incompatible 
measurements and what are they useful for?
 
\item There are other concepts to grasp the nature of measurement in quantum
mechanics, such as coexistence and unbiasedness. What are their applications
in quantum information processing?

\item Some incompatible measurements do not provide an advantage in QRACs, 
so is there a stronger form of incompatibility that is necessary and sufficient 
for QRACs?

\item Another open problem is to clarify the role of incompatible measurements in quantum
metrology, where measurements are used to characterize one or more parameters
of a quantum states with high precision. 

\item One open direction is to investigate the incompatibility properties of quantum measurements that act on many particles, such as on two or more qubits. In general, it would be interesting to characterise the measurement resources that are needed for generating non-classical effects in quantum networks.

 
\end{itemize}
In addition, with the progress of experimental techniques, complex measurements 
with interesting incompatibility features may also become an available resource 
in practical implementations. This will finally lead to further applications of
the theory presented in this review.

\section{Acknowledgements}

We would like to thank  
Alastair Abbott,
Konstantin Beyer,
Nicolas Brunner,
Costantino Budroni,
Tom Bullock,
Paul Busch,
Sébastien Designolle, 
Teiko Heinosaari,
Benjamin D.M. Jones,
Jukka Kiukas,
Matthias Kleinmann, 
Pekka Lahti,
Fabiano Lever,
Kimmo Luoma,
Nikolai Miklin,
Takayuki Miyadera,
Tobias Moroder,
H. Chau Nguyen,
Micha{\l} Oszmaniec,
Marco Piani,
Martin Pl\'avala,
Jussi Schultz,
Michal Sedl\'{a}k,
Jiangwei Shang,
Paul Skrzypczyk,
Walter T. Strunz,
Ivan \v{S}upi\'c,
Armin Tavakoli,
Giuseppe Vitagliano,
Reinhard F. Werner,
Zhen-Peng Xu, 
Kari Ylinen, 
Xiao-Dong Yu,
and
Mario Ziman
for collaborations and discussions on the topic. We are also thankful for the useful comments given by Sébastien Designolle, Huan-Yu Ku, Peter Morgan, Martin Pl\'avala, Marco Túlio Quintino, 
and three anonymous referees on an earlier version of the manuscript.

This work was supported by the Deutsche Forschungsgemeinschaft 
(DFG, German Research Foundation, project numbers 447948357 and 
440958198), the Sino-German Center for Research Promotion (Project M-0294), 
the ERC (Consolidator Grant 683107/TempoQ), the DAAD, the Austrian Science Fund (FWF) P 32273-N27 and the Swiss National Science Foundation (NCCR SwissMAP and Ambizione PZ00P2-202179). 
This research is
supported by the National Research Foundation, Singapore
and A*STAR under its CQT bridging grant.

\bibliography{references}

\end{document}